\documentclass[
 reprint,
 amsmath,amssymb,
 prl,
superscriptaddress
]{revtex4-2}

\usepackage{graphicx}
\usepackage{dcolumn}
\usepackage{bm}
\usepackage{enumitem}
\usepackage{braket}
\usepackage{graphicx}
\usepackage{amsmath}
\usepackage{mathrsfs}
\usepackage{xcolor}
\usepackage{longtable}
\usepackage{amsfonts}
\usepackage{tikz}
\usepackage[colorlinks,linkcolor=blue,citecolor=blue,anchorcolor=blue]{hyperref}
\usepackage{CJK}
\usepackage{upgreek}
\usepackage{array}
\usepackage{mathtools}
\usepackage[T1]{fontenc}

\usepackage{amsmath}
\usepackage[normalem]{ulem}

\newcommand{\dd}[0]{{\ensuremath{d}}}

\begin{document}

\title{Counter-factual carving exponentially improves entangled-state fidelity}

\author{Joshua Ramette}
\affiliation{Department of Physics, MIT-Harvard Center for Ultracold Atoms and Research Laboratory of Electronics, Massachusetts Institute of Technology, Cambridge, Massachusetts 02139, USA}
\author{Josiah Sinclair}
\affiliation{Department of Physics, MIT-Harvard Center for Ultracold Atoms and Research Laboratory of Electronics, Massachusetts Institute of Technology, Cambridge, Massachusetts 02139, USA}
\author{Vladan Vuleti\'c}
\affiliation{Department of Physics, MIT-Harvard Center for Ultracold Atoms and Research Laboratory of Electronics, Massachusetts Institute of Technology, Cambridge, Massachusetts 02139, USA}


\begin{abstract}
We propose a new method, ``counter-factual'' carving, that uses the ``no-jump'' evolution of a probe to generate entangled many-body states of high fidelity.
The probe is coupled to a target ensemble of qubits and engineered to exponentially decay at a rate depending on the target collective spin, such that post-selecting on observing no probe decay precisely removes select faster-decaying spin components.
When probe and $N$-qubit target interact via a cavity mode of cooperativity $C$, counter-factual carving generates entangled states with infidelities of $e^{-C/N}$, an exponential improvement over previous carving schemes.
Counter-factual carving can generate complex entangled states for applications in quantum metrology and quantum computing.
\end{abstract}

\maketitle

Entangled many-body quantum states represent a valuable resource, enabling measurements beyond the standard quantum limit, secure communication networks, and quantum computation \cite{kitagawa1993, duan2001, terhal2015, raussendorf2001, raussendorf2003, Duan2005, Kessler2014, Robinson2024, Malia2022, Sorensen2002, Ramette2022, Reiserer2015}.
To generate an entangled state starting from an easily prepared product state of many qubits, one can either apply a deterministic unitary operation \cite{zhang2023}, or attempt to alter the state vector via a projective measurement onto a subspace of interest \cite{vonneumann_1955}.

Considering an ensemble of $N$ qubits coupled to a cavity, ``carving'' out certain  collective spin components (Dicke states \cite{dicke1954}) $\ket{m}$ via a cavity measurement can prepare the system in a highly entangled state \cite{McConnell2015}.
Such a method was proposed in \cite{chen2015,davis2018} (and realized with similar methods in \cite{Welte2017, Dordevic2021}) where the detection of a multi-frequency photon transmitted through an optical cavity projects the spin system into an entangled superposition of Dicke states, $\sum c_m \ket{m}$.
We call this process “factual” carving, since post-selection occurs on the detection of a transmitted probe photon, which is correlated with the occupation of certain Dicke states of the ensemble.
For factual carving, the infidelity $\epsilon_\textrm{f}$ of the carved state scales as $\epsilon_\textrm{f} \sim (C/N)^{-1}$, where 
$C$ is the cavity cooperativity, characterizing the coupling of a single atom to a single intracavity photon. This infidelity $\epsilon_\textrm{f}$ arises from the finite spectral overlap of the atom-shifted cavity resonances (i.e. different Dicke states) \cite{Suzuki2011, chen2015}.

In this Letter, we propose a new method, which we call ``counter-factual'' carving, that instead is heralded by the absence of the evolution (``no quantum jump'' \cite{carmichael_open_1993, steinberg_quantum_2014}) of a probe coupled to the atomic ensemble.
We find that this method yields exponentially improved infidelity compared to factual carving, scaling as $e^{-C/N}$. The improved scaling arises from engineering an exponential decay which carves away particular spin components.
This method enables the generation of large highly entangled states, for example $N$-atom GHZ states \cite{omran2019,song2019, moses2023}, that are useful for many quantum applications \cite{degen2017,philipp2018, Kessler2014, Li2015}.

\begin{figure}[h]
\includegraphics[width=8.6cm]{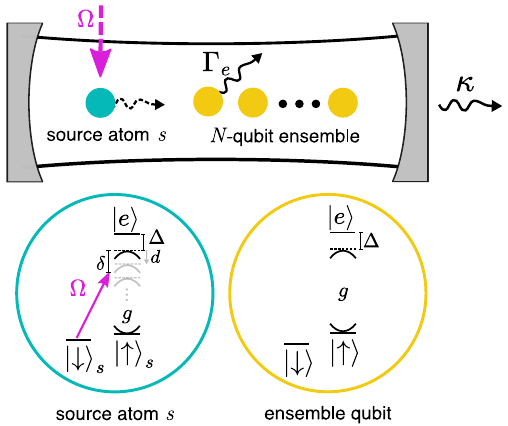}
\caption{Counter-factual carving of a qubit ensemble (target) using an optical cavity.
The cavity contains $N$ qubits (yellow) and a single photon source atom $s$ (teal).
A transverse laser beam $\Omega$ (purple) addresses $s$ and can be frequency-tuned so that the emission on the Raman transition $\ket{\downarrow} \rightarrow \ket{e} \rightarrow \ket{\downarrow}$ in the source atom is tuned to near the cavity resonance (shown in light gray), that is being shifted depending on the collective state of the target qubits (see text). The absence of emission into the cavity, as detectable via the state of the source atom, modifies the target collective state.
}
\label{fig:cavity}
\end{figure}

To demonstrate the principles of counterfactual carving and directly relate it to the most pertinent prior works, we consider a setup similar to \cite{chen2015,davis2018} and shown in Fig.~\ref{fig:cavity}, consisting of an ensemble of $N$ target atoms coupled to a cavity mode, with the addition of a photon source atom $s$ that we use to probe the target.
The internal structure of each atom constitutes a $\Lambda$ system where two ground states $\ket{\uparrow},\ket{\downarrow}$ are coupled to an excited state $\ket{e}$, and for simplicity we assume that $\ket{e}$ decays predominantly to the ground state $\ket{\uparrow}$, and is only weakly coupled to $\ket{\downarrow}$. 
The optical transition $\ket{\uparrow} \rightarrow \ket{e}$ of frequency $\omega_e$ and population decay rate $\Gamma_e$ is coupled to the cavity mode $\hat{a}$ of frequency $\omega_c$ with single-atom Rabi frequency $2g$ and population decay rate $\kappa \ll \Gamma_e$ with a large detuning $\Delta \equiv \omega_e - \omega_c$, with cooperativity $2C \equiv 2 g^2/(\kappa \Gamma_e)$ \cite{kimble1998}

\begin{figure}[h]
\includegraphics[width=8.6cm]{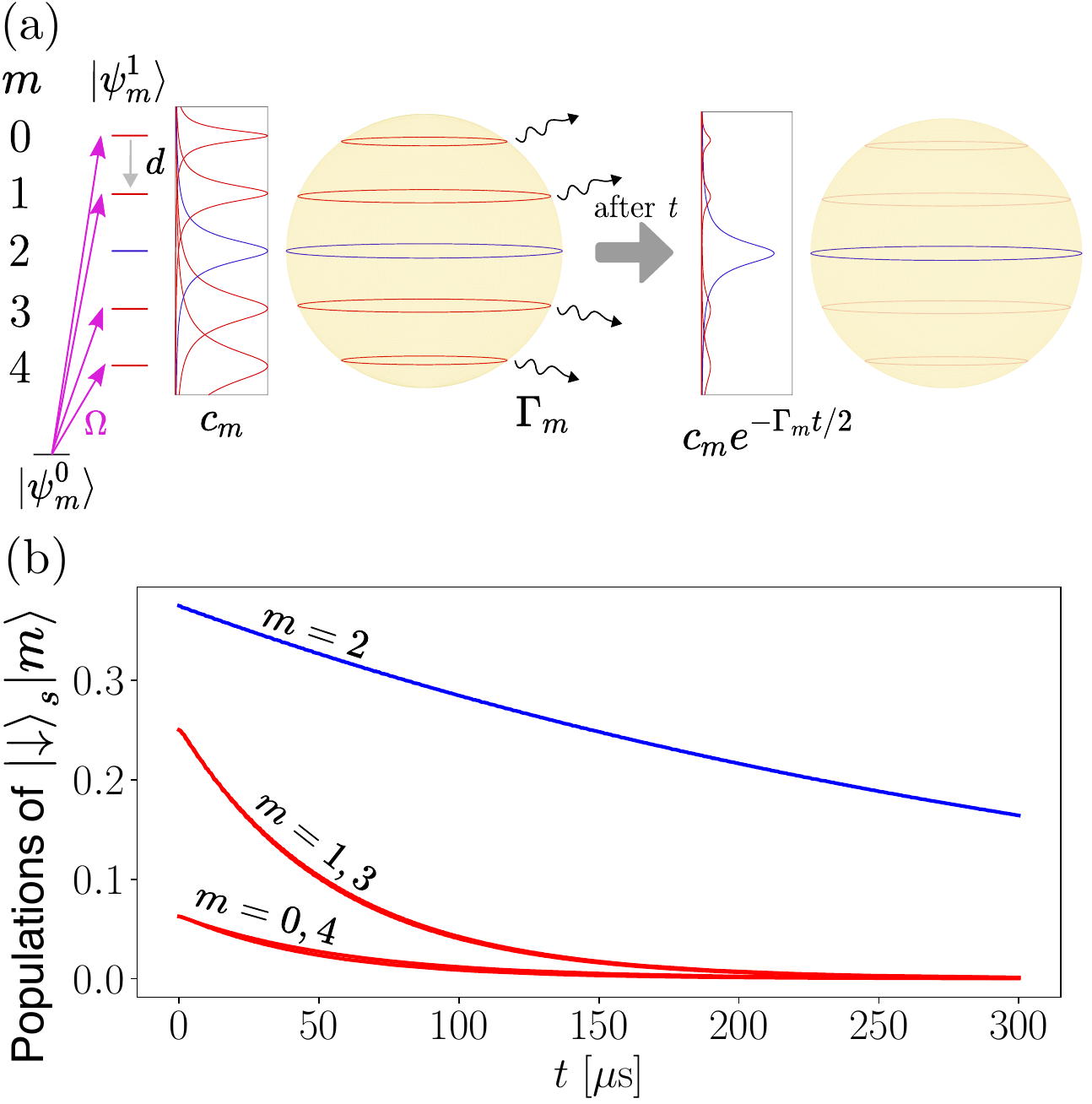}
\caption{
Preparation of the $\ket{m = 2}$ Dicke state by counter-factual carving from a CSS $\propto ( \ket{\uparrow} + \ket{\downarrow} )^{\otimes 4}$ of $N = 4$ atoms. Resonantly addressed levels (red) decay quickly, and the unaddressed level (blue), driven only off-resonantly, decays more slowly.
After time $t$, the populations of the undesired components are exponentially suppressed.
(a)
The ensemble-cavity coupling shifts the cavity mode of width $\kappa$ by $m\dd$.
A tone of $\Omega$ driving $s$ induces a coupling $w = \Omega g/\Delta$ between dressed states $\ket{\psi^0_m}$ and $\ket{\psi^1_m}$.
(b)
Numerical simulation of the master equation for the no-jump evolution of the joint source-target system, showing the remaining populations of $\ket{\downarrow}_s \ket{m}$.
Parameters used were $\kappa = 2 \pi \times 0.2$ MHz, $\Gamma_e = 2 \pi \times 6$ MHz, $g = 2 \pi \times 8.5$ MHz, with $C = 60$, $\Delta = \sqrt{2} g \sqrt{\Gamma_e / \kappa} = 2 \pi \times 66$ MHz.
\label{fig:SchemeDicke}}
\end{figure}

The interaction of the target ensemble with the cavity mode is governed by the Hamiltonian:
\begin{align}
    H_\textrm{target} = -\Delta \hat{a}^\dag \hat{a} + g \hat{a} \sum_{i = 1}^N \ket{e}_i \prescript{}{i}{\bra{\uparrow}} + \textrm{h.c.} \nonumber \\
    = -\Delta \hat{a}^\dag \hat{a} + g \hat{a} \sum_{m = 0}^{N} \sqrt{m} \ket{m_e} \bra{m} + \textrm{h.c.}
    \label{H_ensemble}
\end{align}
where we have re-expressed the Hamiltonian as acting on the collective spin states $\ket{m}$ ($\ket{m}$ is the symmetric state with $m$ atoms in $\ket{\uparrow}$) and $N-m$ atoms in $\ket{\downarrow}$). $\ket{m_e}$ represents the state where $\ket{m}$ has collectively absorbed a photon from the cavity (see Supplement).
This Hamiltonian couples the state $\ket{1}_c \ket{m}$ to $\ket{0}_c \ket{m_e}$ with coupling strength $\sqrt{m}g$ and a detuning of $\Delta$, shifting $\ket{1}_c \ket{m}$ by $m \dd \equiv m g^2/\Delta$ in the dispersive limit $g \ll \Delta$, where assuming $\kappa \ll \Gamma_e$, it is favorable to be far detuned from the excited state.
Here $\ket{p}_c$, with $p=0,1$ denotes the cavity state with $p$ photons.
For sufficiently large single-atom cooperativity $C$, each of these Dicke states, with $m$ atoms in $\ket{\uparrow}$, corresponds to a spectrally resolved Lorentzian line as in Fig.~\ref{fig:SchemeDicke}a.

To probe the $m$-dependent energy shifts and counterfactually carve the desired superposition of Dicke states $\ket{m}$, we use the separately addressable atom $s$, which acts as a single-photon ``source''. (For another use of such a source atom, see \cite{Borregaard2015}.) We initialize $s$ in $\ket{\downarrow}_s$, then couple it to $\ket{e}_s$ with a laser beam of strength $\Omega$ and detuning $\delta$ relative to the empty-cavity resonance.
For simplicity we take the coupling $\Omega$ to be much smaller than all other energy scales in the problem.
The Hamiltonian for $s$ is the same as for the target atoms, but with the additional laser coupling $\Omega$:
\begin{align}
    H_s = - (\Delta & + \delta) \ket{\downarrow}_s \prescript{}{s}{\bra{\downarrow}} \nonumber \\
    & +  \Omega \ket{e}_s \prescript{}{s}{\bra{\downarrow}} + g \hat{a} \ket{e}_s \prescript{}{s}{\bra{\uparrow}} + \textrm{h.c.}
\end{align}
Here the subscript $s$ indicates the source atom, and the total Hamiltonian is $H_\textrm{tot} = H_\textrm{target} + H_s$.

As Fig.~\ref{fig:cavity} illustrates, tuning $\delta$ into resonance with the cavity shifted by a particular Dicke state $\ket{m}$ then enables $s$ to emit a photon into the cavity via the Raman transition $\ket{\downarrow}_s \rightarrow \ket{e}_s \rightarrow \ket{\uparrow}_s$. The photon subsequently leaves the cavity at rate $\kappa$.
In this way the (potential) decay of the source atom ($\ket{\downarrow}_s \rightarrow \ket{\uparrow}_s$) via the cavity reveals the occupation of the corresponding ensemble Dicke state $\ket{m}$.
If the source atom is observed to have not decayed for a time long compared to the characteristic decay time via this channel, then the corresponding amplitude of $\ket{m}$ is exponentially suppressed as in Fig.~\ref{fig:SchemeDicke}.

Diagonalizing $H_\textrm{tot}$ to first order with respect to the atom-cavity coupling, we obtain the dressed states $\ket{\psi^0_m} \equiv \ket{0}_c \ket{\downarrow}_s \ket{m}$ and $\ket{\psi^1_m} \equiv \ket{1}_c \ket{\uparrow}_s \ket{m} - \frac{g}{\Delta} \ket{0}_c \big( \ket{e}_s \ket{m} + \sqrt{m} \ket{\uparrow}_s \ket{m_e} \big)$, where the three components of $\ket{\psi^1_m}$ represent the photon in the cavity, the excited source atom, and the absorption of the photon by the ensemble, respectively.
The state $\ket{\psi^1_m}$ has a decay rate $\kappa_m \equiv \kappa + (m+1)\frac{g^2}{\Delta^2}\Gamma$ due to both the cavity decay ($\kappa$) and scattering from the admixed atomic excited states ($\Gamma_e$) of the source and the $m$ atoms of the target ensemble, where any decay of $\ket{\psi^1_m}$ would leave $s$ in $\ket{\uparrow}_s$.

Then, turning on a single weak tone $\Omega > 0$ with a detuning $\delta$ matching the energy of $\ket{\psi^1_m}$  perturbatively couples $\ket{\psi^0_m}$ to $\ket{\psi^1_m}$ with strength $w \equiv \bra{\psi^1_m} \Omega \ket{e}_s \prescript{}{s}{\bra{\downarrow}}\ket{\psi^0_m} = \Omega g / \Delta$.
We can then write an effective Hamiltonian in terms of the dressed states:
\begin{align}
    H = \sum_{m = 0}^{N} & \biggl[ - \delta \ket{\psi_m^0} \bra{\psi_m^0} - (m+1) d \ket{\psi_m^1} \bra{\psi_m^1} \nonumber \\ &   + w \ket{\psi_m^1} \bra{\psi_m^0} + \textrm{h.c.} \biggr]
    \label{H_dressed}
\end{align}

Here, $w$ couples $\ket{\psi^0_m}$ to state $\ket{\psi^1_m}$, via which it decays into the continuum at a rate \cite{cohen1998}:
\begin{equation}
\Gamma_m = \frac{w^2}{(\kappa_m/2)^2 + (\delta - (m+1)\dd)^2} \kappa_m.
\label{Gamma}
\end{equation}

We denote the quasi-continuum state that $\ket{\psi^0_m}$ decays into as $\ket{0}_c \ket{\uparrow}_s 
\ket{\textrm{L}_m (t)}$, with $\ket{\textrm{L}_m (t) }$ a superposition across states of the ensemble and modes of the environment $E$, with the photon having leaked out of the cavity or having been scattered by an atom.
The initial state with the ensemble in $\ket{m}$ and the environment in the vacuum decays toward this quasi-continuum of scattered photon states, evolving as \cite{scully1997, lukin2016}:
\begin{align}
e^{-\Gamma_m t/2} \ket{0}_c & \ket{\downarrow}_s \ket{m}\ket{\textrm{vac}}_E \nonumber \\
+ & \sqrt{1 - e^{-\Gamma_m t}} \ket{0}_c \ket{\uparrow}_s\ket{\textrm{L}_m(t)},
\label{single_decay}
\end{align}
where we neglect for now any additional overall phase on the first term due to a Stark shift. 

We illustrate the carving procedure by first considering a superposition $\frac{1}{\sqrt{2}}  (\ket{n} + \ket{n+1})$ of just two neighboring Dicke states spaced by $\dd$ in energy. Here $\ket{n+1}$ denotes the Dicke state which we wish to retain after carving, and $\ket{n}$ the state that we strive to annihilate.

To annihilate $\ket{n}$, we tune the  coupling laser $\Omega$ to resonance with the dressed state energy, $\delta = -(n+1)\dd$, so that $\ket{n}$ decays at rate $\Gamma_n = 4w^2/\kappa_n$.
$\Omega$ also addresses $\ket{n+1}$, but off-resonantly by $\dd$, resulting in a slower decay
$\Gamma_{n+1} = \Gamma_n/(1+(2d/\kappa_n)^2)$
(assuming $\kappa_{n+1} \approx \kappa_n$ for optimally chosen value of $\Delta$, as explained below).
The components $\ket{n}, \ket{n+1}$ then evolve according to Eq.~\ref{single_decay}, so that postselecting counter-factually on measuring the source atom after time $t$ to have remained in the state $\ket{\downarrow}_s$ projects the system to the (not normalized) state:
\begin{align}
\ket{\downarrow}_s \big(  e^{-\Gamma_n t/2}  \ket{n} 
 + e^{-\Gamma_{n+1}  t/2} \ket{n+1} \big)
\label{postselected}
\end{align}
Choosing $t=1/\Gamma_{n+1}$, the population of $\ket{n+1}$ decays only by a factor $e^{-1}$, so we have a success probability of $e^{-1}/2$, while the population of $\ket{n}$ is suppressed exponentially, even for modestly large $\dd/\kappa_n$, by a factor $e^{-\Gamma_n/\Gamma_{n+1}}=e^{-(1+(2\dd/\kappa_n)^2)} \gg 1$.

The suppression factor is maximized when 
$\ket{n}$ and $\ket{n+1}$ are maximally distinguishable, i.e. when the atomic detuning $\Delta$ is chosen such that losses between cavity and atomic decay are balanced, $\kappa = n \frac{g^2}{\Delta^2} \Gamma_e$, in which case $\kappa_n = 2 \kappa$ (again, recall we take $\kappa \ll \Gamma_e$). The cooperativity
$C$ then determines the distiguishability since $\dd = \kappa\sqrt{C/ n}$, making $4\dd^2/\kappa_n^2 = C/n$.
The infidelity $\epsilon_\textrm{cf}$ of the state from Eq.~\ref{postselected} with respect to $\ket{n+1}$ is then:
\begin{equation}
    \epsilon_\textrm{cf} = \frac{e^{-C/n}}{1 + e^{-C/n}} \approx e^{-C/n}
    \label{cf_scale}
\end{equation}
for $C/n \gg 1$.
Note the dependence on the ratio $C/n$, where $n$ enters since admixing the excited states of more target qubits further broadens the dressed state linewidth.
This means we can more easily carve away adjacent levels around lower $n$, with $n= 1$ corresponding to a $W$ state, and $n \sim N/2$ for carving states near the equator of the many-atom Bloch sphere, as in Fig.~\ref{fig:SchemeDicke}.

\begin{figure}[h]
\includegraphics[width=8.6cm]{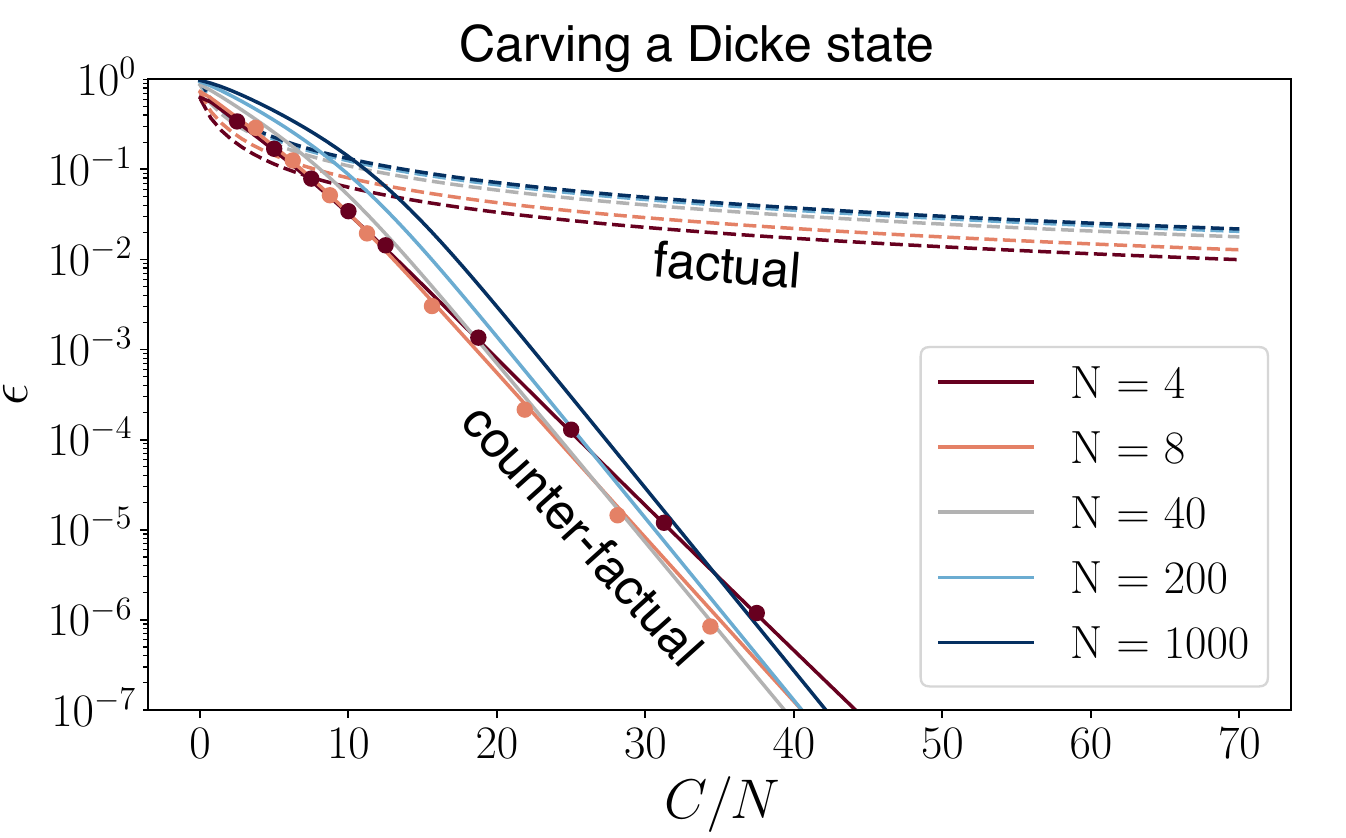}
\caption{
Infidelity $\epsilon$ of carving Dicke states on the equator of the $N$-atom Bloch sphere.
Analytical expressions for factual (dashed) \cite{chen2015} and counter-factual (solid) carving, analogous to Eqs.~\ref{f_scale} and \ref{cf_scale}.
For counter-factual carving, we drive with multiple tones as in Fig.~\ref{fig:SchemeDicke} to annihilate all Dicke levels except $\ket{m = N/2}$.
Solid dots are numerical simulations of counter-factual carving with the master equation, modelling the joint source-cavity-ensemble system for $N = 4$ and 8.
The errors $\epsilon$ shown correspond to similar success probabilities $|c_{N/2}|^2/4$ and $|c_{N/2}|^2/e$ for the factual and counter-factual carving, where the initial population overlap with $\ket{m = N/2}$ is $|c_{N/2}|^2 \approx 1/\sqrt{N}$.
}
\label{fig:dicke_result}
\end{figure}

Having illustrated the principle of counter-factual carving with two neighboring levels $\ket{n}$, $\ket{n+1}$, in Fig.~\ref{fig:dicke_result} we show results when counter-factually carving a highly entangled Dicke state $\ket{m = N/2}$ from an initial coherent spin state (CSS) along the equator of the Bloch sphere, for the same setup as in Fig.~\ref{fig:SchemeDicke}.
The joint source-cavity-ensemble system is prepared in the pure state $\ket{\downarrow}_s \ket{0}_c\ket{+}^{\otimes N}$, where $\ket{+} \equiv (\ket{\downarrow}+\ket{\uparrow})/\sqrt{2}$, and evolves under the total Hamiltonian with simultaneous driving of tones $\Omega$ applied to all levels except $\ket{m = N/2}$.
Analytical expressions for the many-body infidelity $\epsilon$ are obtained by summing contributions $\Gamma_m$ across all driving tones to determine the relative populations of the levels $\ket{m}$.
Each datapoint (large dots) in Fig. \ref{fig:dicke_result} is obtained separately by a simulation of the master equation for the joint source-cavity-ensemble system as in Fig.~\ref{fig:SchemeDicke}b, where the evolution time $t_{1/e}$ is chosen to be when the population of $\ket{\downarrow}_s$ has been reduced by a factor $1/e$.
Then to post-select on measuring no probe evolution, we project the density matrix of the full system $\rho_\textrm{tot}(t = t_{1/e})$ onto $\ket{\downarrow}_s$ and renormalize. 
Finally, we trace out the source atom and the cavity mode, leaving just the reduced density matrix of the ensemble $\rho_\textrm{ensemble}(t = t_{1/e})$, computing its infidelity compared to the Dicke state at the equator $\ket{m=N/2}$, $\epsilon = 1 - \bra{m=N/2} \rho_\textrm{ensemble}(t = t_{1/e})\ket{m=N/2}$.
Fitting the asymptotic behavior, for $N = 8$ we have $\epsilon_\textrm{cf} \approx 1.9 \times e^{-0.41 C/N}$.

\begin{figure}[h]
\includegraphics[width=8.6cm]{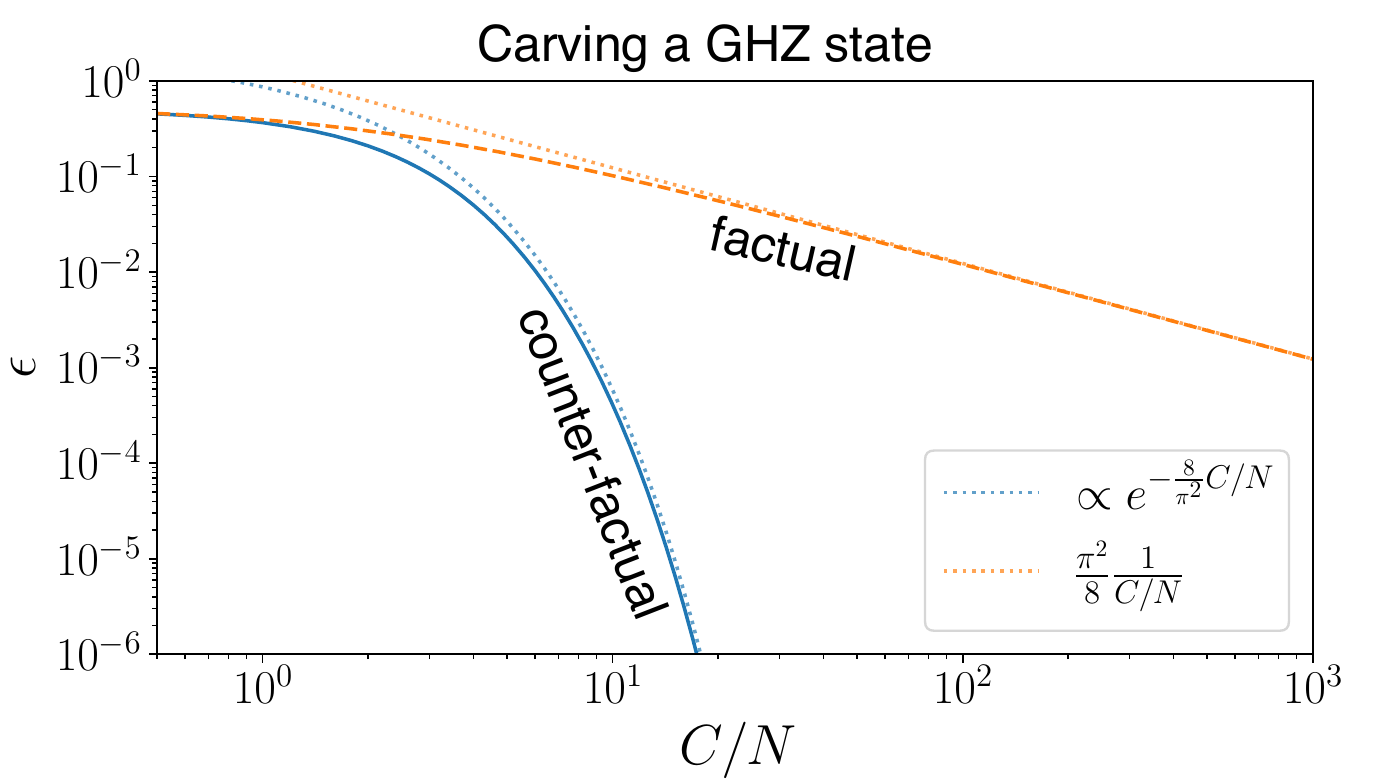}
\caption{
Analytical expressions (see Supplement) showing GHZ state infidelity for a large-$N$ system for counter-factual (blue) and factual (yellow) methods.
Dotted lines indicate asymptotic analytical expressions valid for $C/N \gg 1$.
}
\label{fig:GHZ_result}
\end{figure}

It is also possible to counter-factually carve a superposition of Dicke levels, which we illustrate in Fig.~\ref{fig:GHZ_result} by carving a GHZ state.
This state can be produced
by removing all odd-$m$ Dicke components $\ket{m}$ from the CSS $\ket{+}^{\otimes N}$.
For large $N$, $\ket{+}^{\otimes N}$ is a superposition of Dicke state near the equator, $m \sim N/2$, where the Lorentzians have a (here approximated to be identical) width $\kappa_m \sim 2 \kappa$.
To remove the odd-$m$ states, we apply a resonant tone to each, and for simplicity, we imagine an infinite ladder of such tones resonant with the odd $m$, so that there are no net phase shifts, while all the even-$m$ states decay at identical rates.
In Fig.~\ref{fig:dicke_result} we plot the residual error for factual and counterfactual carving, summing over contributions from all the tones (see Supplement). The result is the same asymptotic scaling as Eq.~\ref{cf_scale}, just with a slightly different numerical coefficient, in this case $\epsilon_\textrm{cf} \approx e^{- \frac{C}{N} \frac{8}{\pi^2}}$ for counterfactual carving, and for comparison, $\epsilon_\textrm{f} \approx \frac{\pi^2}{8} \frac{1}{C/N}$ for the factual method.

The results from Eq.~\ref{cf_scale} and Figs.~\ref{fig:dicke_result}, \ref{fig:GHZ_result} represent an exponential improvement in the scaling of the residual error $\epsilon$ compared to prior ``factual'' carving methods. The infidelity of factual carving schemes can be understood as follows:
Using our source atom $s$, factual carving with a tone tuned to $\ket{n}$ would involve detecting a photon successfully exiting through the cavity mirror, thereby post-selecting on $\ket{\uparrow}_s$ terms where the photon has leaked out of the cavity (as opposed to the $\ket{\downarrow}_s$ terms for counterfactual carving) from Eq.~\ref{single_decay}.
The terms remaining after postselection from Eq.~\ref{single_decay} we can approximate for small $t$, at which point the carving will be of highest fidelity:
\begin{align}
  \sqrt{1 - e^{-\Gamma_{n} t}} \ket{n} + \sqrt{1 - e^{-\Gamma_{n+1} t} }\ket{n+1}\nonumber \\
  \approx \sqrt{\Gamma_{n} t} \ket{n} + \sqrt{\Gamma_{n+1} t }\ket{n+1}
  \label{factual_superposition}
\end{align}

Because factual carving actively drives both $\ket{n}$ and $\ket{n+1}$ toward states with an expelled photon, we can only achieve population distortions proportional to the rates.
The infidelity $\epsilon_\textrm{f}$ for the factual method when attempting to create the state $\ket{n}$ is then:

\begin{equation}
 \epsilon_\textrm{f} = \frac{\Gamma_{n+1}}{\Gamma_{n} + \Gamma_{n+1}} = \frac{1 }{2 + C/n } \xrightarrow[ \frac{C}{n} \gg 1 ]{} \frac{1}{C/n}
  \label{f_scale}
\end{equation}

The same result holds for the photon transmission variant of factual carving \cite{chen2015} due to the polynomial tail of Lorentzian transmission lineshape overlapping with neighboring Dicke states (see Eq.~10 of Ref. \cite{chen2015} and the Supplement).

Counterfactual carving may be used not only to alter the magnitudes of the initial coefficients $c_m$, but also engineer their quantum phases. This can be accomplished by detuning the various frequency component of the drive $\Omega$ slightly from the energies of the corresponding levels $\ket{m}$, such that state-dependent phases are imprinted via the Stark shift. In this way one can carve states with arbitrary phase,
\begin{align}
    \sum_m c_m \ket{m} \rightarrow \sum_m c_m e^{i \phi_m} e^{-\ell_m/2} \ket{m}
    \label{eq:phase_amps}
\end{align}
Imprinting a phase $\phi_m$ as in Eq. \ref{eq:phase_amps} also leads to a small but correctable additional decay for a given level $\ket{m}$.
A tone detuned from a level $m$ by $\delta_m$ for time $t$ results in a phase $\phi = \frac{w^2 \delta_m}{(\kappa_m/2)^2 + \delta_m^2} t$, while the amplitude decays by a factor $e^{-\ell/2}$ with $\ell = \frac{w^2 \kappa_m}{(\kappa_m/2)^2 + \delta_m^2} t$.
By choosing $\delta_m \gtrsim \kappa_m$ it is possible to create an arbitrary phase up to $\pm \pi$ that is accompanied by only a modest decrease in success probability.

Due to the same Stark shift, Dicke states other than the one that is being addressed also experience unwanted phase shifts. 
In the Supplement, we outline an iterative method, which takes into account how the phase and amplitude control pulses affect neighboring levels $m$.
As each pulse $\Omega$ near a given level $m$ disturbs the phases and amplitudes of its neighbors by a known amount smaller in $C/N$, this disturbance can be fed forward, correcting the phases and amplitude disturbances caused by neighbors to rapidly converge to the desired state with only a modest overall decrease in success probability.
The ability to modify the phases and amplitudes in this fashion allows for the correction of a host of additional small effects one could consider, including for example experimental imperfections and approximations made in our description of the GHZ state carving.
With the ability to control both amplitudes and phases of different Dicke states, we now have access to a toolbox for carving high fidelity many-body states while requiring only moderate cooperativity $C$.

In addition to achieving exponentially better fidelity, counter-factual carving is of fundamental interest as it harnesses the ``no jump'' \cite{carmichael_open_1993, steinberg_quantum_2014} evolution of a probe coupled to a qubit ensemble to exploit a curious property of quantum measurements, how merely giving the probe the possibility to evolve is sufficient to alter the quantum state even in instances where no probe evolution is observed.
Counter-factual carving also has the advantage that it automatically leaves the probe in a tensor product with the ensemble qubits, unlike the factual method where the probe must be carefully engineered to evolve in precisely the same way for all components, lest it become entangled differently with each component and lead to decoherence.

We note that counter-factual carving could also be heralded by post-selecting on measuring a photon to be expelled at a late time after the addressed levels have decayed away, but this seemingly factual post-selection is simply another way to measure that the probe-ensemble system underwent the same no-jump evolution we describe up to that point in time.

While in the above analysis, for the sake of definiteness, we have applied counter-factual carving in the setting of cavity QED, relating it to the most pertinent prior works \cite{chen2015,davis2018}, counter-factual carving is a general approach applicable to many quantum systems.
We have included in the Supplement a more general treatment which may be helpful for applying our approach in other contexts.

In summary, we have proposed a method that can create complex many-body entangled states.
Errors are suppressed exponentially in the cooperativity, a parameter that quantifies the ratio of coherent coupling to decoherence of the system.

\section{Acknowledgement}

J.R. invented the counter-factual carving method, worked out the scaling properties, and drafted the manuscript. J.R. and J.S. together conceptually developed the work and the intuitive understanding of carving presented in this manuscript. V.V. supervised the project. All authors discussed the method and results, and contributed to the manuscript.

\section{Supplement}

\subsection{Hamiltonian}

We can obtain the form of the Hamiltonian we use in the main text, starting from just the terms for the cavity mode, each atom in the ensemble, and the direct couplings between them:
\begin{align}
    H_\textrm{target} = \omega_c \hat{a}^\dag \hat{a} + \sum_{i = 1}^N \omega_e \ket{e}_i \prescript{}{i}{\bra{e}} + g \hat{a} \ket{e}_i \prescript{}{i}{\bra{\uparrow}} +  \textrm{h.c.} \nonumber \\
    = (n_e + \hat{a}^\dag \hat{a}) \omega_e -\Delta \hat{a}^\dag \hat{a} + \sum_{i = 1}^N g \hat{a} \ket{e}_i \prescript{}{i}{\bra{\uparrow}} + \textrm{h.c.} \nonumber \\
\end{align}
where $\Delta \equiv \omega_e - \omega_c$ and $n_e \equiv \sum_{i = 1}^N \ket{e}_i \prescript{}{i}{\bra{e}}$ is the number of target atoms in the excited state.
We can also drop the term $(n_e + \hat{a}^\dag \hat{a})$, since we work in the subspace with one total excitation, which can only be exchanged by the coupling terms between a photonic or atomic excitation.

Additionally, since we consider the ensemble initialized in a Dicke state $\ket{m}$ or a superposition of such states, we just need to analyze how each $\ket{m}$ for $m = 0,...,N$ is coupled to other states by the Hamiltonian.
Since each term in $\ket{m}$ has $m$ atoms in $\ket{\uparrow}$, $H_\textrm{target}$ maps each to a state with $m$ terms, with a different atom excited to $\ket{e}$ on each.
We can then define the properly renormalized state $\ket{m_e}$, the state where $\ket{m}$ has collectively absorbed a photon from the cavity, shared among all the atoms that were in $\ket{\uparrow}$ and coupled to the cavity, such that $\sqrt{m} \ket{m_e} \equiv \big(\sum_{i = 1}^N \ket{e}_i \prescript{}{i}{\bra{\uparrow}} \big) \ket{m}$.
We can then re-express the Hamiltonian as a $\sqrt{m}$-enhanced coupling between the states $\ket{m}$ and $\ket{m_e}$:
\begin{align}
    H_\textrm{target} = -\Delta \hat{a}^\dag \hat{a} + g \hat{a} \sum_{m = 0}^{N} \sqrt{m} \ket{m_e} \bra{m} + \textrm{h.c.}
    \label{H_ensemble}
\end{align}

This Hamiltonian couples the state $\ket{1}_c \ket{m}$ to $\ket{0}_c \ket{m_e}$ with coupling strength $\sqrt{m}g$ and a detuning of $\Delta$, shifting $\ket{1}_c \ket{m}$ by $m \dd \equiv m g^2/\Delta$ in the dispersive limit $g \ll \Delta$. ($\ket{p}_c$, with $p=0,1$ denotes the cavity state with $p$ photons.)

$H_\textrm{target}$ is then the full Hamiltonian for the target ensemble, and we can then add the source atom $s$, which has the same atom-cavity coupling term as each ensemble atom, but now also with an energy term for $\ket{\downarrow}_s$ and a coupling between $\ket{\downarrow}_s$ and $\ket{e}_s$ driven by the laser $\Omega$ detuned by $\Delta + \delta$ below the excited state of $s$:
\begin{align}
    H_s = - (\Delta & + \delta) \ket{\downarrow}_s \prescript{}{s}{\bra{\downarrow}} \nonumber \\
    & +  \Omega \ket{e}_s \prescript{}{s}{\bra{\downarrow}} + g \hat{a} \ket{e}_s \prescript{}{s}{\bra{\uparrow}} + \textrm{h.c.}
\end{align}
The total Hamiltonian is $H_\textrm{tot} = H_\textrm{target} + H_s$.

Additionally, each excited state has decay rate $\Gamma_e$ (decaying to $\ket{\uparrow}$; this branching can be made nearly perfect in practical settings \cite{Borregaard2015}), while the cavity mode has decay rate $\kappa$.
Any decay then permanently removes the system from the 1-excitation subspace due to the loss of a photon to the environment, making the simulation of the master equation straightforward, as we never subsequently drive $\ket{\uparrow}$.

\subsection{General Model}
\begin{figure}[h]
\includegraphics[width=8.6cm]{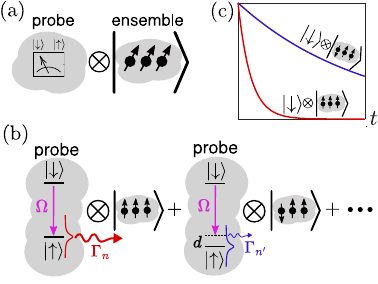}
\caption{A general model of counter-factual carving. (a) Two-level probe and an atomic ensemble in a CSS.
(b) A perturbative drive $\Omega$ rotates the probe $\ket{\downarrow} \rightarrow \ket{\uparrow}$.
The probe state $\ket{\uparrow}$, shifted depending on $\ket{m}$, decays to $\ket{\downarrow}$ more quickly when resonant.
(c) Components with larger $\Gamma$ decay more quickly.
Modest differences in decay rates translate into exponentially large ratios of final state amplitudes after time $t$.
}
\label{probe_ensemble}
\end{figure}

For a general model of counterfactual carving, we consider as in Fig.~\ref{probe_ensemble}a a probe and a qubit ensemble, engineering the interaction so that the no-jump evolution of the probe translates to carving of the ensemble states.
Connecting to our discussion within cavity QED in the main text, the probe states here $\ket{\uparrow}, \ket{\downarrow}$ correspond to the source atom and cavity mode states $\ket{0}_c \ket{\downarrow}_s$, $\ket{1}_c \ket{\uparrow}_s$.

We initialize the probe in $\ket{\downarrow}$ and the ensemble in some CSS.
To carve and herald entanglement, we need to induce an interaction between the probe and ensemble, to rotate the probe from $\ket{\downarrow}$ to $\ket{\uparrow}$ conditional on the ensemble being in a certain collective spin state $\ket{m}$.
This could be done, as in Fig.~\ref{probe_ensemble}b, by a direct coupling, introducing a term $\alpha_m \ket{\uparrow}\bra{\uparrow} \otimes \ket{m} \bra{m}$, where the energy level $\ket{\uparrow}$ of the probe is shifted in a way dependent on $m$, with a characteristic energy spacing of order  $d$  between adjacent levels.
In the main text, this interaction is realized by coupling the ensemble and the probe to the cavity mode bus.
Then, weakly driving the probe with a drive of strength $\Omega$ and frequency $\omega$ resonant with the shift $\alpha_{n}$ will make the probe rotate $\ket{\downarrow}\ket{n} \rightarrow \ket{\uparrow}\ket{n}$ if the ensemble is in $\ket{n}$, but remain $\ket{\downarrow}\ket{n'} \rightarrow \ket{\downarrow}\ket{n'}$ if it is in $\ket{n'}$, so that subsequently measuring the probe in $\ket{\downarrow}$ removes $\ket{n}$ from the CSS.

In general, however, we expect any system, such as the cavity mode, capable of simultaneously coupling to many qubits to also inevitably couple to unwanted environmental degrees of freedom (which we can think of as some quasicontinuum), causing $\ket{\uparrow}$ to acquire some linewidth $\kappa$. 
In the case of a cavity mode mediating a coupling between a source atom and a qubit ensemble, this linewidth would be caused by the decay of the cavity photon mediating the interaction.
This broadening of $\ket{\uparrow}$ prohibits perfectly resolving individual levels $m$, making it impossible to rotate some components without touching others.

This limitation turns out to be fundamental for both factual and counterfactual carving. In both cases,
the weak drive $\Omega$ will then actually make the probe ``decay'' from $\ket{\downarrow}$ through the state $\ket{\uparrow}$ to some state  $\ket{\textrm{Loss}_m}$ in a way that could depend on $\ket{m}$.
Here letting $\delta_m \equiv \omega - \alpha_m$ be the detuning between the probe drive and the energy of $\ket{m}$, with $\Omega$ on each level will decay out of the state $\ket{\downarrow}$ with a rate \cite{cohen1998}:
\begin{equation}
\Gamma_m(\delta_m) = \Omega^2 \frac{\kappa}{(\kappa/2)^2 + \delta_m^2}
\label{Gamma_general}
\end{equation}
where a level addressed resonantly decays at a rate $\Gamma_m(0) = \Omega^2 \frac{4}{\kappa}$, whereas a level addressed far off resonant decays at a slower rate of $\Gamma_{m}(\delta_m \gg \kappa) \rightarrow \Omega^2 \frac{\kappa}{\delta_m^2}$.
When directly addressing a level $\ket{n}$, a nearby level shifted by $d$ in energy will then decay at a rate that is a factor $\frac{4 \kappa^2}{d^2}$ slower.
Due to the broadening of $\ket{\uparrow}$, the probe will evolve $\ket{\downarrow} \rightarrow \ket{\textrm{Loss}_m}$ at least a small amount for all $m$, limiting the selectivity with which we can make the evolution of the probe state depend only on a particular level $\ket{m}$, and thus limiting our ability to perfectly carve levels $m$ out of the ensemble state. 

However, by carving counter-factually we use this decay mechanism to achieve an exponentially large contrast between the populations of $\ket{n}$ and $\ket{n'}$ (see Fig.~\ref{probe_ensemble}c).
As in Fig.~\ref{probe_ensemble}a, we initialize the probe in the state $\ket{\downarrow}$ and the ensemble into some CSS. We separate the CSS out into spin components, and illustrate by isolating the component $\ket{n'}$, which we would like to retain after carving, and a component $\ket{n}$, which we would like to annihilate:
\begin{equation}
    \ket{\downarrow} \sum_m c_m \ket{m} =  c_{n}\ket{\downarrow} \ket{n} + c_{n'} \ket{\downarrow} \ket{n'} + ...
\end{equation}
Then, turning on $\Omega$ as in Fig.~\ref{probe_ensemble}b, resonant with the probe transition when the ensemble is in state $\ket{n}$, leads to the decay of the probe at a different rate for each level:
\begin{align}
c_n\big( & e^{-\Gamma_n t/2} \ket{\downarrow} + \sqrt{1 - e^{-\Gamma_n t} } \ket{\textrm{Loss}_n} \big) \ket{n}\label{many_terms_general} \\
 &+ c_{n'} \big( e^{-\Gamma_{n'} t/2} \ket{\downarrow} + \sqrt{1 - e^{-\Gamma_{n'} t} } \ket{\textrm{Loss}_{n'}} \big) \ket{n'} + \dotsb \notag
\end{align}
Postselecting on measuring the probe to have remained in the state $\ket{\downarrow}$ projects the system to the unnormalized state:
\begin{equation}
\ket{\downarrow} \big(  c_n e^{-\Gamma_n t/2}  \ket{n} 
 + c_{n'} e^{-\Gamma_{n'}  t/2} \ket{n'} + \dotsb \big)
\label{postselected_general}
\end{equation}
where if $\Gamma_n$ is modestly larger than $\Gamma_n'$, $\ket{n}$ has been exponentially suppressed compared to $\ket{n'}$, as in Fig.~\ref{probe_ensemble}c.
Allowing Eq.~\ref{postselected_general} to evolve for a time $t = 1/\Gamma_{n'}$, the population of $\ket{n'}$ has decayed to $1/e$, but the population of $\ket{n}$ has been suppressed exponentially to $e^{-\Gamma_{n}/ \Gamma_{n'}} = e^{-4 d^2/\kappa^2}$.

In contrast, factual carving would involve post-selecting on finding the probe to have evolved out of $\ket{\downarrow}$, where the resonantly addressed level $\ket{n}$ is now the level we wish to project the ensemble into.
Postselecting in this way, the terms remaining from Eq.~\ref{many_terms_general} we can approximate for small $t$, at which point the carving will be of highest fidelity:
\begin{align}
    & c_n \sqrt{1 - e^{-\Gamma_{n} t}} \ket{\textrm{Loss}_n}\ket{n} + c_{n'}\sqrt{1 - e^{-\Gamma_{n'} t} }\ket{\textrm{Loss}_{n'}}\ket{n'} + \dotsb \nonumber \\
    &\approx c_n \sqrt{\Gamma_{n} t} \ket{\textrm{Loss}_n} \ket{n} + c_{n'}\sqrt{\Gamma_{n'} t}\ket{\textrm{Loss}_{n'}}\ket{n'} + \dotsb
    \label{factual_superposition}
\end{align}
The fidelity of the factual method relies on the probe moving into the state $\ket{\uparrow}$ faster for $\ket{n}$ than $\ket{n'}$, but the population of the component $\ket{n'}$ we want to annihilate is now only suppressed polynomially in $d/\kappa$, by a factor $\Gamma_{n'}/\Gamma_n = 1/(4 d^2/\kappa^2)$.
Because with factual carving we drive the probe into $\ket{\uparrow}$ by a polynomial amount whether the ensemble is in $\ket{n}$ or in $\ket{n'}$, the best we can do is achieve population distortions proportional to the rates.
Furthermore, note that when generalizing to carve using multiple tones of $\Omega$ to leave a superposition of several desired levels, the state $\ket{\textrm{Loss}_m}$ of the probe can become entangled with the various levels $m$, which would need to be somehow erased to maintain a factorizable superposition of levels of the ensemble.

\subsection{Factual Carving Infidelity $\epsilon_\textrm{f}$}

From \cite{chen2015} using an incident photon for factual carving, the Lorentzian transmission spectrum looks like:
\begin{equation}
    \mathcal{T}(\delta,m) = \frac{1}{1 + \frac{4 m C}{1 + 4(\Delta + \delta)^2/\Gamma_e^2} - 2i \Big[ \frac{\delta}{\kappa} - 4mC \frac{(\delta+\Delta)/\Gamma_e}{1 + 4 (\Delta + \delta)^2/\Gamma_e^2}\Big]} \nonumber
\end{equation}

We assume $\Delta \gg \Gamma, \delta$, which simplifies to:
\begin{equation}
    \mathcal{T}(\delta,m) = \frac{1}{1 + m C \frac{\Gamma_e^2}{\Delta^2} - 2i \Big[ \frac{\delta}{\kappa} - m C \frac{\Gamma_e}{\Delta} \Big] }
\end{equation}

To maximally distinguish the Lorentzians around some particular level $m$, we optimize the performance by adjusting $\Delta$ such that $\kappa = m \frac{g^2}{\Delta^2} \Gamma_e$, or $\Delta^2 = m C \Gamma_e^2$.
Also using $mC\frac{\Gamma}{\Delta} = m \frac{g^2}{\kappa \Delta} = m \frac{d}{\kappa}$, we can express:
\begin{equation}
    \mathcal{T}(\delta,m) = \frac{1}{2} \frac{1}{1 - i \big[ \frac{\delta - m d}{\kappa} \big] }
\end{equation}
where it is clear that tuning $\delta$ onto resonance with $md$ maximizes the transmission.

Now to carve between two different Dicke levels, $\ket{n}$ and $\ket{n+1}$ (with $\Delta$ chosen to optimize around the level $n$), we tune $\delta$ to $nd$ and look at the resulting transmission contrast for the two states.
For $\ket{n}$ we get $\mathcal{T}(\delta = nd, n) = \frac{1}{2}$, and for $\ket{n+1}$ we get:
\begin{equation}
    \mathcal{T}(\delta = nh, n+1) = \frac{1}{2} \frac{1}{1 - i \big[ \frac{h}{\kappa} \big] } = \frac{1}{2} \frac{1}{1 - i \sqrt{C/n} }
\end{equation}

Post-selecting on detecting a transmitted photon then projects an initial superposition of $\ket{n}$ and $\ket{n+1}$ onto the following unnormalized state, distorted by the relative transmission amplitudes:
\begin{multline}
\ket{n} + \ket{n+1} \rightarrow \\
\mathcal{T}(\delta = nd, n) \ket{n} + \mathcal{T}(\delta = nd, n+1) \ket{n+1} \nonumber \\
= \frac{1}{2}(\ket{n} + \frac{1}{1 - i \sqrt{C/n} } \ket{n+1})
\end{multline}

The infidelity $\epsilon_\textrm{f}$ of the relative population of the unwanted state $\ket{n+1}$ is then:
\begin{equation}
    \epsilon_\textrm{f} = \frac{|\mathcal{T}(\delta = nd, n+1)|^2}{|\mathcal{T}(\delta = nd, n)|^2 + |\mathcal{T}(\delta = nd, n+1)|^2}
    \label{epsilon_factual}
\end{equation}
which is:
\begin{equation}
    \epsilon_\textrm{f} = \frac{\frac{1}{1 + C/n} }{1 + \frac{1}{1 + C/n} } = \frac{1 }{2 + C/n }
\end{equation}

\subsection{GHZ State}

We can carve a GHZ state by removing all the $\ket{m}$ with $m$ odd from the CSS $\ket{+}^{\otimes N}$, since the components with an odd number of qubits in $\ket{\downarrow}$ cancel in the superposition $\frac{1}{\sqrt{2}} \big(\ket{+}^{\otimes N} + \ket{-}^{\otimes N} \big)$.
To remove the odd $m$, we apply a resonant tone to each.
For simplicity, we imagine an infinite ladder of tones of identical strength spaced by $2d$ and resonant with the odd $m$.
For $N$ large, $\ket{+}^{\otimes N}$ has negligible support except near the center of the many-atom Bloch sphere around where $m \sim \frac{N}{2}$, so we set $\Delta$ such that $\kappa = \frac{N}{2} \frac{g^2}{\Delta^2} \Gamma_e$ and then approximate $\kappa_m = \kappa + m \frac{g^2}{\Delta^2} \Gamma_e \approx 2 \kappa$ for $m \approx N/2$ where the CSS is centered around the equator of the Bloch sphere.

Note that for $N$ not asymptotically large, the Lorentzians across the support of the CSS will have slightly larger widths for larger $m$, so the even states we want to preserve will be decaying at slightly different rates above and below.
We can approximately fix this with a pulse that swaps $\ket{\uparrow}$ and $\ket{\downarrow}$ halfway through the counterfactual carve, swapping $m$ with $N - m$.
The level sees the same infinite ladder of tones before and after the flip, so we can look at how each tone detuned from $m$ by $\delta_m$ affects the level both before and after the flip, experiencing a total decay $e^{-\Gamma_m t/2} e^{-\Gamma_{N-m} t/2}$.
This is approximately independent of $m$ since $\Gamma_m \approx \frac{w^2}{\delta_m^2} \kappa_m$, giving $\Gamma_m t /2 + \Gamma_{N-m} t/2 = \Gamma_{N/2} t$ by using $\kappa_m$ from the previous paragraph.
Imperfections associated with these assumptions can be further corrected if necessary using the techniques from the next section on iterative phase control.

Turning on this ladder of tones resonant with the odd $m$ levels, each level sees no net phase shift, and recalling Eq.~\ref{Gamma}, each level $m$ now decays with a total decay rate $\Gamma^\textrm{tot}_m$ due to contributions from multiple driving tones $\Omega$, with distinct values $\Gamma^\textrm{tot}_\textrm{even}$ and $\Gamma^\textrm{tot}_\textrm{odd}$ for even and odd $m$.
The odd levels decay at a rate given by the resonant tone and by off resonant tones above and below in energy detuned by multiples of $2d$, with $d = \kappa \sqrt{C/ m}$ when optimized with $m \sim N/2$. Letting $\Gamma_\textrm{res} \equiv \frac{2 w^2}{\kappa}$ be the decay rate due to a resonant tone, the total decay rate for an odd level is:
\begin{align}
    \Gamma^\textrm{tot}_{\textrm{odd}} = \Gamma_\textrm{res} + \Gamma_\textrm{res} \times 2 \sum_{j=1}^\infty \frac{1}{1 + (2j)^2 \times C/ (N/2)}  \nonumber \\
    \xrightarrow[ \frac{C}{N} \gg 1 ]{} \Gamma_\textrm{res} \Big(1 + \frac{\pi^2}{24}  \frac{1}{C/N} \Big)
\end{align}

Similarly, for even $m$, which we wish to preserve, are addressed off resonantly by tones above and below detuned by a distance of $d$, $3d$, $5d$, ..., giving:
\begin{align}
    \Gamma^\textrm{tot}_{\textrm{even}} = \Gamma_\textrm{res} \times 2 \sum_{j} \frac{1}{1 + (2j-1)^2 \times C/ (N/2)} \nonumber \\
    \xrightarrow[ \frac{C}{N} \gg 1 ]{} \Gamma_\textrm{res} \frac{\pi^2}{8} \frac{1}{C/N}
\end{align}

With these two distinct values for the decay rates of the even and odd $m$ levels, we can separate our initial CSS state into even and odd components, with $\ket{\textrm{even } m} \equiv \frac{1}{\sqrt{2}} \big(\ket{+}^{\otimes N} + \ket{-}^{\otimes N} \big)$ and $\ket{\textrm{odd } m} \equiv \frac{1}{\sqrt{2}} \big(\ket{+}^{\otimes N} - \ket{-}^{\otimes N} \big)$, then evolves similarly to Eq.~\ref{single_decay} as:
\begin{align}
    \ket{+}^{\otimes N} = & \frac{1}{\sqrt{2}} (\ket{\textrm{even } m} + \ket{\textrm{odd } m}) \nonumber \\ 
    \rightarrow \frac{1}{\sqrt{2}} & (e^{-\Gamma^\textrm{tot}_{\textrm{even}} t/2} \ket{\textrm{even } m} + e^{-\Gamma^\textrm{tot}_{\textrm{odd}}t/2} \ket{\textrm{odd } m} ) \nonumber \\
    &+ \textrm{ photon  scattering terms}
    \label{even_odd_evolve}
\end{align}
where we have made $\Gamma^\textrm{tot}_{\textrm{odd}} > \Gamma^\textrm{tot}_{\textrm{even}}$ by tuning laser tones into resonance with the odd $m$.

To carve a GHZ state factually, we can proceed as in Eq.~\ref{factual_superposition}, preparing $\ket{\textrm{odd } m}$ by evolving Eq.~\ref{even_odd_evolve} for a small $t$, after which with $\mathcal{O}(1)$ probability we have a residual population error $\epsilon_\textrm{f} = \Gamma^\textrm{tot}_{\textrm{even}}/(\Gamma^\textrm{tot}_{\textrm{odd}} + \Gamma^\textrm{tot}_{\textrm{even}})$, or:
\begin{align}
    \epsilon_\textrm{f} = \frac{\Gamma^\textrm{tot}_{\textrm{even}}/\Gamma^\textrm{tot}_{\textrm{odd}}}{\Gamma^\textrm{tot}_{\textrm{even}}/\Gamma^\textrm{tot}_{\textrm{odd}} + 1} \xrightarrow[ \frac{C}{N} \gg 1 ]{} \frac{\pi^2}{8} \frac{1}{C/N}
    \label{GHZ_f_error}
\end{align}

To carve the GHZ state $\ket{\textrm{even } m}$ counter-factually, we choose $t = 1/\Gamma^\textrm{tot}_{\textrm{even}}$ so that the even levels will have decayed by only an amount $1/e$, guaranteeing a success probability of at least $\frac{1}{2} \frac{1}{e}$. Post-selecting on the terms from Eq.~\ref{even_odd_evolve} with no photon scattering, we have $\epsilon_\textrm{cf} = e^{-\Gamma^\textrm{tot}_{\textrm{odd}}t}/( e^{-\Gamma^\textrm{tot}_{\textrm{odd}}t} + e^{-\Gamma^\textrm{tot}_{\textrm{even}}t})$.
The residual error is then an ``exponentiated'' form of Eq.~\ref{GHZ_f_error}:
\begin{align}
    \epsilon_\textrm{cf} = & \frac{e^{-\Gamma^\textrm{tot}_{\textrm{odd}}/\Gamma^\textrm{tot}_{\textrm{even}}}}{e^{-\Gamma^\textrm{tot}_{\textrm{odd}} /\Gamma^\textrm{tot}_{\textrm{even}} } + e^{-1}} \nonumber \\
    & \xrightarrow[ \frac{C}{N} \gg 1 ]{} e \times e^{-\Gamma^\textrm{tot}_{\textrm{odd}}/\Gamma^\textrm{tot}_{\textrm{even}} } = e^{\frac{2}{3}} \times e^{-\frac{8}{\pi^2} C/N }
    \label{GHZ_cf_error}
\end{align}

We plot Eqs.~\ref{GHZ_f_error}-\ref{GHZ_cf_error} in Fig.~\ref{fig:GHZ_result} of the main text, showing both the full expression summed over all tones converging to these asymptotic scaling formulas for large $C/N$.
Note that the time $t$ as well as $C/N$ both scale as $\ln(1/\epsilon_\textrm{cf})$.

\subsection{Phase Control}

Here, we outline an iterative method to counterfactually adjust both phases and amplitudes on the many-body state. 

While the amplitude reduction of a Dicke level by addressing it resonantly is discussed in the main text, by detuning the drive $\Omega$ slightly from the energy of $\ket{m}$, we can induce a Stark shift on $\ket{m}$ which allows us to imprint a phase.
Adjusting amplitude and detuning for potentially multiple tones, we can engineer a set of phases $\{ \phi_m \}$ and a set of amplitudes $\{ \ell_m \}$ onto the many-body state:

\begin{align}
    \sum_m c_m \ket{m} \rightarrow \sum_m c_m e^{i \phi_m} e^{-\ell_m/2} \ket{m}
    \label{phase_amps_sup}
\end{align}

Consider a system with large $C/N$, so that we have a relatively well distinguished Lorentzian feature for each $m$.
We can iterate over the $m$, detuning a beam slightly from the energy of each state $\ket{m}$ and applying each of the desired phases $\{ \phi_m \}$. 
We can then also iterate across all the $m$ and similarly target each with a resonant beam to apply the desired amplitude reductions $\{ \ell_m \}$.
After the first such round applying such phases and amplitudes, the state will be nearly Eq.~\ref{phase_amps_sup}, up to a small correction arising from two effects, 1) the unwanted amplitude disturbance which occurs when a phase is applied, and 2) the fact that pulses targeting a given level $m$ will have disturbed the phases and amplitudes of its neighbors.

We will show that the first effect can be dealt with and that the second effect occurs as a higher order in $N/C \ll 1$. Consequently, in the subsequent round we can feed forward a correction where laser pulses iterate once more over all the $m$, correcting the small phases and amplitude disturbances remaining from the previous round.
Each successive round, the required correction becomes higher in order of $N/C$, and we can therefore rapidly converge to the desired state (Eq.~\ref{phase_amps_sup}) with only a modest overall decrease in success probability.
The ability to touch up the phases and amplitudes in this fashion allows for the correction of a host of additional small effects one could consider, including for example experimental imperfections and approximations made in our description of the GHZ carve.

\subsubsection{Iterative Method Make Errors Exponentially Converge}

To show that the iterative method will cause the remaining unwanted phase and amplitude disturbances to rapidly converge to zero, we first place a bound on the magnitude of unwanted phase and amplitude disturbances on the $T+1$ round in terms of those from the prior round and show that they are of a different order in $N/C$.

If a tone is applied resonantly to a level $m$, it produces a reduction in amplitude $\ell_m$ and no phase distortion arises on $m$, but other levels do experience both amplitude and phase disturbances at higher orders of $N/C$.
To imprint a phase $\phi_m$ on level $m$, a perturbative tone is instead detuned from level $m$ by $\delta_m$ for time $t$, resulting in the phase: 
\begin{equation}
    \phi_m = \frac{w^2 \delta_m t}{(\kappa_m/2)^2 + \delta_m^2}
\end{equation}
If this tone is detuned $b$ linewidths $\kappa_m$ away from $m$, so $\delta_m = b \kappa_m$, there will be unwanted amplitude decay
\begin{equation}
    \frac{w^2 \kappa_m t}{(\kappa_m/2)^2 + \delta_m^2} = \phi_m/b,
\end{equation}
implying only a modest decrease in success probability $e^{-\phi_m/b}$ to apply an arbitrary phase up to $\pm \pi$.
Assuming we can detune by a few linewidths ($\delta_m \gg \kappa_m$), and the Lorentzians are spaced by much more than this ($d \gg \delta_m$), then we can approximate $\phi_m = \frac{w^2 t}{\delta_m}$.
Recalling how $d = \sqrt{C/N} \kappa_m/2$ (taking the worst case by optimizing around $m = N$), Table \ref{ell_phi_table} shows how applying a phase shift or an amplitude reduction to level $m$ affects the phases and amplitudes of that level, and also how it affects the phases and amplitudes of a neighboring level $k$.



\newcolumntype{L}{|>{\bfseries}c|} 

\renewcommand{\arraystretch}{2}

\begin{table}
    \centering
    \begin{tabular}{Lc|c|c|c|}
    \hline
    \textbf{} & \multicolumn{2}{c|}{$m$} & \multicolumn{2}{c|}{$k$} \\
    \cline{2-5}
    & $\ell_m'$ & $\phi_m'$ & $\ell_k'$ & $\phi_k'$ \\
    \hline
    $\ell_m$ & $\ell_m$ & 0 & $\frac{\ell_m}{(k-m)^2} \frac{N}{C}$ & $\frac{\ell_m}{2(k-m)} \sqrt{\frac{ N}{C }}$ \\
    $\phi_m$ & $\frac{\phi_m}{b}$ & $\phi_m$ & $\frac{4 \phi_m b}{(k-m)^2} \frac{N}{C}$ & $\frac{2 \phi_m b}{(k-m)} \sqrt{ \frac{ N}{C }}$\\
    \hline
    \end{tabular}
    \caption{How applying $\ell_m$ or $\phi_m$ to a level $m$ affects the phase and amplitude of that same level as well as that of a neighboring level $k$.
    $\ell_m$ is applied via a resonant pulse to the level $m$, and $\phi_m$ is applied via a detuned pulse to the level $m$.
    Primed variables represent how those pulses affect the phase and amplitude of the level $m$ itself, as well as a neighboring level $k$, which is an energy spacing of $(k-m)d$ away.
    }
    \label{ell_phi_table}
\end{table}

We will break down our pulses into $T$ rounds, and on the first round, $T=0$, we simply apply our target phases and amplitude pulses as $\{ \phi_m^{(0)} \}$ and $\{ \ell_m^{(0)} \}$, the phases and amplitudes we want to have on the final state we carve.
After applying these, however there will be corrections at a higher order in $N/C$ that we will have to correct on an additional round.
On the following round $T+1$, we can then apply pulses to each $m$ to exactly reverse the unwanted phase shift from the last round, and we can further reduce the amplitude of each $m$ to even out undesired amplitude disturbances from the previous round.

Consider the sum total of ways a level $m$ on a round of corrections $T$ can have its phase and amplitude affected.
In addition to having had its proper targeted application of phase $\phi_m^{(T)}$ and amplitude reduction $\ell_m^{(T)}$, its amplitude overall may have been reduced in the worst case by a factor of $e^{-\phi_{max}^{(T)}/b}$ from the target phase application (where the maximum would be found by looking at all the levels and considering the one which required the largest phase correction this round). We can correct this distortion by applying a ``leveling'' pulse, reducing each by the amount required to make the amplitudes even. After this all $m$ levels will be reduced by a factor of at most $e^{-\phi_{max}^{(T)}/b}$.

Unfortunately, for each of its neighbors $k$, these same steps (applying the desired phase $\phi_k^{(T)}$ and amplitude $\ell_k^{(T)}$ adjustments and then the leveling adjustment $\phi_{max}^{(T)}/b$) results in an unwanted phase disturbance of up to:
\begin{align}
     \phi_{dist}^{(T)} \leq \sqrt{N/C} \frac{2b \phi_k^{(T)} + (\ell_k^{(T)} + \phi_{max}^{(T)}/b)/2 }{k-m}
\end{align}
and an unwanted amplitude disturbance of
\begin{align}
    \ell_{dist}^{(T)} \leq (N/C) \frac{4 b \phi_k^{(T)} + (\ell_k^{(T)} + \phi_{max}^{(T)}/b)}{(k-m)^2},
\end{align}
where each contribution is taken directly from Table \ref{ell_phi_table}: the first term is due to the neighbor's phase shift, and the term in parentheses comes from the neighbor's direct amplitude reduction as well as applying the possible leveling correction, an amplitude reduction smaller than $\phi_{max}^{(T)}/b$.

Summing over all $k$ and using $\sum_{k=0}^N \frac{1}{(k-m)} \leq  2(1 + \ln{N}) \equiv A$ and $\sum_{k=0}^N \frac{1}{(k-m)^2} \leq  2 \times \frac{\pi^2}{6} \equiv B$, the resulting total unwanted phase and amplitude disturbance that occur during round $T$ on level $m$ due to all other levels $k$ can be simplified and bounded:
\begin{align}
    \phi_{max}^{(T+1)} \leq \sqrt{N/C} \sum_k \frac{2 b \phi_k^{(T)} + (\ell_k^{(T)} + \phi_{max}^{(T)}/b)/2 }{k-m} \\
    \leq \sqrt{N/C} A (2b \phi_{max}^{(T)} + \phi_{max}^{(T)}/(2b) + \ell_{max}^{(T)}/4)
\end{align}
\begin{align}
    \ell_{max}^{(T+1)} \leq (N/C) \sum_k \frac{4 b \phi_k^{(T)} + (\ell_k^{(T)} + \phi_{max}^{(T)}/b) }{(k-m)^2} \\
    \leq (N/C) B (4b \phi_{max}^{(T)} + \phi_{max}^{(T)}/b + \ell_{max}^{(T)}),
\end{align}
where $\phi_{max}^{T+1}$ and $\ell_{max}^{T+1}$ here are the largest unwanted phase and amplitude disturbances occurring from pulses on neighboring levels on the round $T$, and are therefore the largest values that we will have to apply in the next round $T+1$ as our target corrections.
We note that $A$ grows with $N$, but only logarithmically in the worst-case bound we present here (where phases are all conspiring to add completely constructively).

Next, let $x \equiv \textrm{max}\{\sqrt{N/C} A (2b + 1/(2b)), (N/C) B (4b + 1/b) \}$.  
Summing the expressions for $\phi_{max}^{T+1}$ and $\ell_{max}^{T+1}$ from the previous paragraph, we then have the following inequalities:
\begin{align}
    \phi_{max}^{(T+1)} + \ell_{max}^{(T+1)}
    \leq (2x) \left[ \phi_{max}^{(T)} + \ell_{max}^{(T)} \right] \\
    \leq (2x)^{T+1} \left[ \phi_{max}^{(0)} + \ell_{max}^{(0)} \right].
\end{align}
For $(2x) < 1$, meaning for sufficiently large $C/N$, the remaining phase and amplitude disturbances $\phi_{max}^{(T+1)} + \ell_{max}^{(T+1)}$ decrease exponentially with the number of rounds of correction $T$.

\subsubsection{Iterative Method Minimally Affects Success Probability}

Next, we show that the multiple rounds of phase and amplitude corrections minimally decreases the overall success probability.
For a given level $m$ on round $T$ of correction, the amount of unwanted population reduction due to applying phases $\phi_m^{(T)}$ is at most $e^{-\phi_{max}^{(T)}/b}$, and the maximum amount of population reduction happening to any given level $m$ due to the addressing of neighboring levels $k$ is $e^{-\ell_{max}^{(T+1)}}$ (recall this quantity is the maximum amount of population disturbance of any of the $m$ levels due to neighboring pulses during the round $T$, and hence the subsequent maximum amplitude correction necessary on round $T+1$).

After $T$ rounds, the success probability has then been reduced by unwanted disturbances by no more than 
\begin{align}
    (e^{-\phi_{max}^{(T=0)}/b} e^{-\ell_{max}^{(T=1)}}) \times (e^{-\phi_{max}^{(T=1)}/b} e^{-\ell_{max}^{(T=2)}}) \times \dots \\
    = e^{-\phi_{max}^{(T=0)}/b} \bigg(\prod_{j=1}^T e^{-\ell_{max}^{(j)}} e^{-\phi_{max}^{(j)}/b} \bigg) e^{-\phi_{max}^{(T+1)}/b}
\end{align}
Then, since $\phi_{max}^{(T)}/b + \ell_{max}^{(T)} \leq \phi_{max}^{(T)} + \ell_{max}^{(T)} \leq (2x)^{T} (\phi_{max}^{(0)} + \ell_{max}^{(0)})$, by the $T$-th round the remaining success probability would have been reduced by no more than a factor:
\begin{align}
   e^{-\phi_{max}^{(T=0)}/b} \bigg( \prod_{j=1}^T  e^{-(\phi_{max}^{(j)} + \ell_{max}^{(j)})} \bigg) e^{-\phi_{max}^{(T+1)}/b} \nonumber\\
=  e^{-\phi_{max}^{(T=0)}/b} e^{-\sum_{j=1}^T (2x)^{j} (\phi_{max}^{(T=0)} + \ell_{max}^{(T=0)})} e^{-\phi_{max}^{(T+1)}/b} \nonumber\\
\xrightarrow[ T \rightarrow \infty ]{} e^{-\phi_{max}^{(T=0)}/b} e^{-\frac{2x}{1-2x} (\phi_{max}^{(T=0)} + \ell_{max}^{(T=0)})}
\label{iterative_prob_reduced}
\end{align}

In the last line of Eq.~\ref{iterative_prob_reduced}, the first factor (worst case $e^{-\pi/b}$) is due to the direct application of initial phases, and the second factor is due to pulses applying phase and amplitude reductions to neighbors.
With $N/C$ (and thereby $x$) sufficiently small so that $\frac{2x}{1-2x} (\phi_{max}^{(T=0)} + \ell_{max}^{(T=0)}) = 1$, we maintain the same constant success probability of $e^{-1}$ we saw in the main text.
Since $\phi_{max}^{(T=0)} \leq \pi$ and because $\ell_{max}^{(T=0)}$ exponentially suppresses unwanted spin components, we only need polynomially small $N/C$ to maintain this condition and with constant success probability carve many-body states with exponentially small infidelities.

These bounds are just to show that arbitrary phases and amplitudes can be applied efficiently even in a worst-case scenario; in a practical setting many unwanted phase distortions would in fact cancel out and one expects.

\bibliography{ramette_biblio}

\begin{thebibliography}{34}%
\makeatletter
\providecommand \@ifxundefined [1]{%
 \@ifx{#1\undefined}
}%
\providecommand \@ifnum [1]{%
 \ifnum #1\expandafter \@firstoftwo
 \else \expandafter \@secondoftwo
 \fi
}%
\providecommand \@ifx [1]{%
 \ifx #1\expandafter \@firstoftwo
 \else \expandafter \@secondoftwo
 \fi
}%
\providecommand \natexlab [1]{#1}%
\providecommand \enquote  [1]{``#1''}%
\providecommand \bibnamefont  [1]{#1}%
\providecommand \bibfnamefont [1]{#1}%
\providecommand \citenamefont [1]{#1}%
\providecommand \href@noop [0]{\@secondoftwo}%
\providecommand \href [0]{\begingroup \@sanitize@url \@href}%
\providecommand \@href[1]{\@@startlink{#1}\@@href}%
\providecommand \@@href[1]{\endgroup#1\@@endlink}%
\providecommand \@sanitize@url [0]{\catcode `\\12\catcode `\$12\catcode `\&12\catcode `\#12\catcode `\^12\catcode `\_12\catcode `\%12\relax}%
\providecommand \@@startlink[1]{}%
\providecommand \@@endlink[0]{}%
\providecommand \url  [0]{\begingroup\@sanitize@url \@url }%
\providecommand \@url [1]{\endgroup\@href {#1}{\urlprefix }}%
\providecommand \urlprefix  [0]{URL }%
\providecommand \Eprint [0]{\href }%
\providecommand \doibase [0]{https://doi.org/}%
\providecommand \selectlanguage [0]{\@gobble}%
\providecommand \bibinfo  [0]{\@secondoftwo}%
\providecommand \bibfield  [0]{\@secondoftwo}%
\providecommand \translation [1]{[#1]}%
\providecommand \BibitemOpen [0]{}%
\providecommand \bibitemStop [0]{}%
\providecommand \bibitemNoStop [0]{.\EOS\space}%
\providecommand \EOS [0]{\spacefactor3000\relax}%
\providecommand \BibitemShut  [1]{\csname bibitem#1\endcsname}%
\let\auto@bib@innerbib\@empty
\bibitem [{\citenamefont {Kitagawa}\ and\ \citenamefont {Ueda}(1993)}]{kitagawa1993}%
  \BibitemOpen
  \bibfield  {author} {\bibinfo {author} {\bibfnamefont {M.}~\bibnamefont {Kitagawa}}\ and\ \bibinfo {author} {\bibfnamefont {M.}~\bibnamefont {Ueda}},\ }\bibfield  {title} {\bibinfo {title} {Squeezed spin states},\ }\href {https://doi.org/10.1103/PhysRevA.47.5138} {\bibfield  {journal} {\bibinfo  {journal} {Phys. Rev. A}\ }\textbf {\bibinfo {volume} {47}},\ \bibinfo {pages} {5138} (\bibinfo {year} {1993})}\BibitemShut {NoStop}%
\bibitem [{\citenamefont {Duan}\ \emph {et~al.}(2001)\citenamefont {Duan}, \citenamefont {Lukin}, \citenamefont {Cirac},\ and\ \citenamefont {Zoller}}]{duan2001}%
  \BibitemOpen
  \bibfield  {author} {\bibinfo {author} {\bibfnamefont {L.-M.}\ \bibnamefont {Duan}}, \bibinfo {author} {\bibfnamefont {M.~D.}\ \bibnamefont {Lukin}}, \bibinfo {author} {\bibfnamefont {J.~I.}\ \bibnamefont {Cirac}},\ and\ \bibinfo {author} {\bibfnamefont {P.}~\bibnamefont {Zoller}},\ }\bibfield  {title} {\bibinfo {title} {Long-distance quantum communication with atomic ensembles and linear optics},\ }\href {https://doi.org/10.1038/35106500} {\bibfield  {journal} {\bibinfo  {journal} {Nature}\ }\textbf {\bibinfo {volume} {414}},\ \bibinfo {pages} {413} (\bibinfo {year} {2001})}\BibitemShut {NoStop}%
\bibitem [{\citenamefont {Terhal}(2015)}]{terhal2015}%
  \BibitemOpen
  \bibfield  {author} {\bibinfo {author} {\bibfnamefont {B.~M.}\ \bibnamefont {Terhal}},\ }\bibfield  {title} {\bibinfo {title} {Quantum error correction for quantum memories},\ }\href {https://doi.org/10.1103/RevModPhys.87.307} {\bibfield  {journal} {\bibinfo  {journal} {Rev. Mod. Phys.}\ }\textbf {\bibinfo {volume} {87}},\ \bibinfo {pages} {307} (\bibinfo {year} {2015})}\BibitemShut {NoStop}%
\bibitem [{\citenamefont {Raussendorf}\ and\ \citenamefont {Briegel}(2001)}]{raussendorf2001}%
  \BibitemOpen
  \bibfield  {author} {\bibinfo {author} {\bibfnamefont {R.}~\bibnamefont {Raussendorf}}\ and\ \bibinfo {author} {\bibfnamefont {H.~J.}\ \bibnamefont {Briegel}},\ }\bibfield  {title} {\bibinfo {title} {A one-way quantum computer},\ }\href {https://doi.org/10.1103/PhysRevLett.86.5188} {\bibfield  {journal} {\bibinfo  {journal} {Phys. Rev. Lett.}\ }\textbf {\bibinfo {volume} {86}},\ \bibinfo {pages} {5188} (\bibinfo {year} {2001})}\BibitemShut {NoStop}%
\bibitem [{\citenamefont {Raussendorf}\ \emph {et~al.}(2003)\citenamefont {Raussendorf}, \citenamefont {Browne},\ and\ \citenamefont {Briegel}}]{raussendorf2003}%
  \BibitemOpen
  \bibfield  {author} {\bibinfo {author} {\bibfnamefont {R.}~\bibnamefont {Raussendorf}}, \bibinfo {author} {\bibfnamefont {D.~E.}\ \bibnamefont {Browne}},\ and\ \bibinfo {author} {\bibfnamefont {H.~J.}\ \bibnamefont {Briegel}},\ }\bibfield  {title} {\bibinfo {title} {Measurement-based quantum computation on cluster states},\ }\href {https://doi.org/10.1103/PhysRevA.68.022312} {\bibfield  {journal} {\bibinfo  {journal} {Phys. Rev. A}\ }\textbf {\bibinfo {volume} {68}},\ \bibinfo {pages} {022312} (\bibinfo {year} {2003})}\BibitemShut {NoStop}%
\bibitem [{\citenamefont {Duan}\ \emph {et~al.}(2005)\citenamefont {Duan}, \citenamefont {Wang},\ and\ \citenamefont {Kimble}}]{Duan2005}%
  \BibitemOpen
  \bibfield  {author} {\bibinfo {author} {\bibfnamefont {L.-M.}\ \bibnamefont {Duan}}, \bibinfo {author} {\bibfnamefont {B.}~\bibnamefont {Wang}},\ and\ \bibinfo {author} {\bibfnamefont {H.~J.}\ \bibnamefont {Kimble}},\ }\bibfield  {title} {\bibinfo {title} {Robust quantum gates on neutral atoms with cavity-assisted photon scattering},\ }\href {https://doi.org/10.1103/PhysRevA.72.032333} {\bibfield  {journal} {\bibinfo  {journal} {Phys. Rev. A}\ }\textbf {\bibinfo {volume} {72}},\ \bibinfo {pages} {032333} (\bibinfo {year} {2005})}\BibitemShut {NoStop}%
\bibitem [{\citenamefont {Kessler}\ \emph {et~al.}(2014)\citenamefont {Kessler}, \citenamefont {K\'om\'ar}, \citenamefont {Bishof}, \citenamefont {Jiang}, \citenamefont {S\o{}rensen}, \citenamefont {Ye},\ and\ \citenamefont {Lukin}}]{Kessler2014}%
  \BibitemOpen
  \bibfield  {author} {\bibinfo {author} {\bibfnamefont {E.~M.}\ \bibnamefont {Kessler}}, \bibinfo {author} {\bibfnamefont {P.}~\bibnamefont {K\'om\'ar}}, \bibinfo {author} {\bibfnamefont {M.}~\bibnamefont {Bishof}}, \bibinfo {author} {\bibfnamefont {L.}~\bibnamefont {Jiang}}, \bibinfo {author} {\bibfnamefont {A.~S.}\ \bibnamefont {S\o{}rensen}}, \bibinfo {author} {\bibfnamefont {J.}~\bibnamefont {Ye}},\ and\ \bibinfo {author} {\bibfnamefont {M.~D.}\ \bibnamefont {Lukin}},\ }\bibfield  {title} {\bibinfo {title} {Heisenberg-limited atom clocks based on entangled qubits},\ }\href {https://doi.org/10.1103/PhysRevLett.112.190403} {\bibfield  {journal} {\bibinfo  {journal} {Phys. Rev. Lett.}\ }\textbf {\bibinfo {volume} {112}},\ \bibinfo {pages} {190403} (\bibinfo {year} {2014})}\BibitemShut {NoStop}%
\bibitem [{\citenamefont {Robinson}\ \emph {et~al.}(2024)\citenamefont {Robinson}, \citenamefont {Miklos}, \citenamefont {Tso}, \citenamefont {Kennedy}, \citenamefont {Bothwell}, \citenamefont {Kedar}, \citenamefont {Thompson},\ and\ \citenamefont {Ye}}]{Robinson2024}%
  \BibitemOpen
  \bibfield  {author} {\bibinfo {author} {\bibfnamefont {J.~M.}\ \bibnamefont {Robinson}}, \bibinfo {author} {\bibfnamefont {M.}~\bibnamefont {Miklos}}, \bibinfo {author} {\bibfnamefont {Y.~M.}\ \bibnamefont {Tso}}, \bibinfo {author} {\bibfnamefont {C.~J.}\ \bibnamefont {Kennedy}}, \bibinfo {author} {\bibfnamefont {T.}~\bibnamefont {Bothwell}}, \bibinfo {author} {\bibfnamefont {D.}~\bibnamefont {Kedar}}, \bibinfo {author} {\bibfnamefont {J.~K.}\ \bibnamefont {Thompson}},\ and\ \bibinfo {author} {\bibfnamefont {J.}~\bibnamefont {Ye}},\ }\bibfield  {title} {\bibinfo {title} {Direct comparison of two spin-squeezed optical clock ensembles at the 10 level},\ }\bibfield  {journal} {\bibinfo  {journal} {Nature Physics}\ }\href {https://doi.org/10.1038/s41567-023-02310-1} {10.1038/s41567-023-02310-1} (\bibinfo {year} {2024})\BibitemShut {NoStop}%
\bibitem [{\citenamefont {Malia}\ \emph {et~al.}(2022)\citenamefont {Malia}, \citenamefont {Wu}, \citenamefont {Mart{\'i}nez-Rinc{\'o}n},\ and\ \citenamefont {Kasevich}}]{Malia2022}%
  \BibitemOpen
  \bibfield  {author} {\bibinfo {author} {\bibfnamefont {B.~K.}\ \bibnamefont {Malia}}, \bibinfo {author} {\bibfnamefont {Y.}~\bibnamefont {Wu}}, \bibinfo {author} {\bibfnamefont {J.}~\bibnamefont {Mart{\'i}nez-Rinc{\'o}n}},\ and\ \bibinfo {author} {\bibfnamefont {M.~A.}\ \bibnamefont {Kasevich}},\ }\bibfield  {title} {\bibinfo {title} {Distributed quantum sensing with mode-entangled spin-squeezed atomic states},\ }\href {https://doi.org/10.1038/s41586-022-05363-z} {\bibfield  {journal} {\bibinfo  {journal} {Nature}\ }\textbf {\bibinfo {volume} {612}},\ \bibinfo {pages} {661} (\bibinfo {year} {2022})}\BibitemShut {NoStop}%
\bibitem [{\citenamefont {S\o{}rensen}\ and\ \citenamefont {M\o{}lmer}(2002)}]{Sorensen2002}%
  \BibitemOpen
  \bibfield  {author} {\bibinfo {author} {\bibfnamefont {A.~S.}\ \bibnamefont {S\o{}rensen}}\ and\ \bibinfo {author} {\bibfnamefont {K.}~\bibnamefont {M\o{}lmer}},\ }\bibfield  {title} {\bibinfo {title} {Entangling atoms in bad cavities},\ }\href {https://doi.org/10.1103/PhysRevA.66.022314} {\bibfield  {journal} {\bibinfo  {journal} {Phys. Rev. A}\ }\textbf {\bibinfo {volume} {66}},\ \bibinfo {pages} {022314} (\bibinfo {year} {2002})}\BibitemShut {NoStop}%
\bibitem [{\citenamefont {Ramette}\ \emph {et~al.}(2022)\citenamefont {Ramette}, \citenamefont {Sinclair}, \citenamefont {Vendeiro}, \citenamefont {Rudelis}, \citenamefont {Cetina},\ and\ \citenamefont {Vuleti\ifmmode~\acute{c}\else \'{c}\fi{}}}]{Ramette2022}%
  \BibitemOpen
  \bibfield  {author} {\bibinfo {author} {\bibfnamefont {J.}~\bibnamefont {Ramette}}, \bibinfo {author} {\bibfnamefont {J.}~\bibnamefont {Sinclair}}, \bibinfo {author} {\bibfnamefont {Z.}~\bibnamefont {Vendeiro}}, \bibinfo {author} {\bibfnamefont {A.}~\bibnamefont {Rudelis}}, \bibinfo {author} {\bibfnamefont {M.}~\bibnamefont {Cetina}},\ and\ \bibinfo {author} {\bibfnamefont {V.}~\bibnamefont {Vuleti\ifmmode~\acute{c}\else \'{c}\fi{}}},\ }\bibfield  {title} {\bibinfo {title} {Any-to-any connected cavity-mediated architecture for quantum computing with trapped ions or rydberg arrays},\ }\href {https://doi.org/10.1103/PRXQuantum.3.010344} {\bibfield  {journal} {\bibinfo  {journal} {PRX Quantum}\ }\textbf {\bibinfo {volume} {3}},\ \bibinfo {pages} {010344} (\bibinfo {year} {2022})}\BibitemShut {NoStop}%
\bibitem [{\citenamefont {Reiserer}\ and\ \citenamefont {Rempe}(2015)}]{Reiserer2015}%
  \BibitemOpen
  \bibfield  {author} {\bibinfo {author} {\bibfnamefont {A.}~\bibnamefont {Reiserer}}\ and\ \bibinfo {author} {\bibfnamefont {G.}~\bibnamefont {Rempe}},\ }\bibfield  {title} {\bibinfo {title} {Cavity-based quantum networks with single atoms and optical photons},\ }\href {https://doi.org/10.1103/RevModPhys.87.1379} {\bibfield  {journal} {\bibinfo  {journal} {Rev. Mod. Phys.}\ }\textbf {\bibinfo {volume} {87}},\ \bibinfo {pages} {1379} (\bibinfo {year} {2015})}\BibitemShut {NoStop}%
\bibitem [{\citenamefont {Zhang}\ \emph {et~al.}(2023)\citenamefont {Zhang}, \citenamefont {Chi},\ and\ \citenamefont {Hu}}]{zhang2023}%
  \BibitemOpen
  \bibfield  {author} {\bibinfo {author} {\bibfnamefont {T.}~\bibnamefont {Zhang}}, \bibinfo {author} {\bibfnamefont {Z.}~\bibnamefont {Chi}},\ and\ \bibinfo {author} {\bibfnamefont {J.}~\bibnamefont {Hu}},\ }\href@noop {} {\bibinfo {title} {Entanglement generation via single-qubit rotations in a teared hilbert space}} (\bibinfo {year} {2023}),\ \Eprint {https://arxiv.org/abs/2312.04507} {arXiv:2312.04507} \BibitemShut {NoStop}%
\bibitem [{\citenamefont {Von~Neumann}(1955)}]{vonneumann_1955}%
  \BibitemOpen
  \bibfield  {author} {\bibinfo {author} {\bibfnamefont {J.}~\bibnamefont {Von~Neumann}},\ }\href@noop {} {\emph {\bibinfo {title} {Mathematical Foundations of Quantum Mechanics}}}\ (\bibinfo  {publisher} {Princeton University Press},\ \bibinfo {year} {1955})\BibitemShut {NoStop}%
\bibitem [{\citenamefont {Dicke}(1954)}]{dicke1954}%
  \BibitemOpen
  \bibfield  {author} {\bibinfo {author} {\bibfnamefont {R.~H.}\ \bibnamefont {Dicke}},\ }\bibfield  {title} {\bibinfo {title} {Coherence in spontaneous radiation processes},\ }\href {https://doi.org/10.1103/PhysRev.93.99} {\bibfield  {journal} {\bibinfo  {journal} {Phys. Rev.}\ }\textbf {\bibinfo {volume} {93}},\ \bibinfo {pages} {99} (\bibinfo {year} {1954})}\BibitemShut {NoStop}%
\bibitem [{\citenamefont {McConnell}\ \emph {et~al.}(2015)\citenamefont {McConnell}, \citenamefont {Zhang}, \citenamefont {Hu}, \citenamefont {Cuk},\ and\ \citenamefont {Vuleti\'c}}]{McConnell2015}%
  \BibitemOpen
  \bibfield  {author} {\bibinfo {author} {\bibfnamefont {R.}~\bibnamefont {McConnell}}, \bibinfo {author} {\bibfnamefont {H.}~\bibnamefont {Zhang}}, \bibinfo {author} {\bibfnamefont {J.}~\bibnamefont {Hu}}, \bibinfo {author} {\bibfnamefont {S.}~\bibnamefont {Cuk}},\ and\ \bibinfo {author} {\bibfnamefont {V.}~\bibnamefont {Vuleti\'c}},\ }\bibfield  {title} {\bibinfo {title} {Entanglement with negative wigner function of almost 3,000 atoms heralded by one photon},\ }\href {https://doi.org/10.1038/nature14293} {\bibfield  {journal} {\bibinfo  {journal} {Nature}\ }\textbf {\bibinfo {volume} {519}},\ \bibinfo {pages} {439} (\bibinfo {year} {2015})}\BibitemShut {NoStop}%
\bibitem [{\citenamefont {Chen}\ \emph {et~al.}(2015)\citenamefont {Chen}, \citenamefont {Hu}, \citenamefont {Duan}, \citenamefont {Braverman}, \citenamefont {Zhang},\ and\ \citenamefont {Vuleti\ifmmode~\acute{c}\else \'{c}\fi{}}}]{chen2015}%
  \BibitemOpen
  \bibfield  {author} {\bibinfo {author} {\bibfnamefont {W.}~\bibnamefont {Chen}}, \bibinfo {author} {\bibfnamefont {J.}~\bibnamefont {Hu}}, \bibinfo {author} {\bibfnamefont {Y.}~\bibnamefont {Duan}}, \bibinfo {author} {\bibfnamefont {B.}~\bibnamefont {Braverman}}, \bibinfo {author} {\bibfnamefont {H.}~\bibnamefont {Zhang}},\ and\ \bibinfo {author} {\bibfnamefont {V.}~\bibnamefont {Vuleti\ifmmode~\acute{c}\else \'{c}\fi{}}},\ }\bibfield  {title} {\bibinfo {title} {Carving complex many-atom entangled states by single-photon detection},\ }\href {https://doi.org/10.1103/PhysRevLett.115.250502} {\bibfield  {journal} {\bibinfo  {journal} {Phys. Rev. Lett.}\ }\textbf {\bibinfo {volume} {115}},\ \bibinfo {pages} {250502} (\bibinfo {year} {2015})}\BibitemShut {NoStop}%
\bibitem [{\citenamefont {Davis}\ \emph {et~al.}(2018)\citenamefont {Davis}, \citenamefont {Wang}, \citenamefont {Safavi-Naeini},\ and\ \citenamefont {Schleier-Smith}}]{davis2018}%
  \BibitemOpen
  \bibfield  {author} {\bibinfo {author} {\bibfnamefont {E.~J.}\ \bibnamefont {Davis}}, \bibinfo {author} {\bibfnamefont {Z.}~\bibnamefont {Wang}}, \bibinfo {author} {\bibfnamefont {A.~H.}\ \bibnamefont {Safavi-Naeini}},\ and\ \bibinfo {author} {\bibfnamefont {M.~H.}\ \bibnamefont {Schleier-Smith}},\ }\bibfield  {title} {\bibinfo {title} {Painting nonclassical states of spin or motion with shaped single photons},\ }\href {https://doi.org/10.1103/PhysRevLett.121.123602} {\bibfield  {journal} {\bibinfo  {journal} {Phys. Rev. Lett.}\ }\textbf {\bibinfo {volume} {121}},\ \bibinfo {pages} {123602} (\bibinfo {year} {2018})}\BibitemShut {NoStop}%
\bibitem [{\citenamefont {Welte}\ \emph {et~al.}(2017)\citenamefont {Welte}, \citenamefont {Hacker}, \citenamefont {Daiss}, \citenamefont {Ritter},\ and\ \citenamefont {Rempe}}]{Welte2017}%
  \BibitemOpen
  \bibfield  {author} {\bibinfo {author} {\bibfnamefont {S.}~\bibnamefont {Welte}}, \bibinfo {author} {\bibfnamefont {B.}~\bibnamefont {Hacker}}, \bibinfo {author} {\bibfnamefont {S.}~\bibnamefont {Daiss}}, \bibinfo {author} {\bibfnamefont {S.}~\bibnamefont {Ritter}},\ and\ \bibinfo {author} {\bibfnamefont {G.}~\bibnamefont {Rempe}},\ }\bibfield  {title} {\bibinfo {title} {Cavity carving of atomic bell states},\ }\href {https://doi.org/10.1103/PhysRevLett.118.210503} {\bibfield  {journal} {\bibinfo  {journal} {Phys. Rev. Lett.}\ }\textbf {\bibinfo {volume} {118}},\ \bibinfo {pages} {210503} (\bibinfo {year} {2017})}\BibitemShut {NoStop}%
\bibitem [{\citenamefont {Ðorđević}\ \emph {et~al.}(2021)\citenamefont {Ðorđević}, \citenamefont {Samutpraphoot}, \citenamefont {Ocola}, \citenamefont {Bernien}, \citenamefont {Grinkemeyer}, \citenamefont {Dimitrova}, \citenamefont {Vuletić},\ and\ \citenamefont {Lukin}}]{Dordevic2021}%
  \BibitemOpen
  \bibfield  {author} {\bibinfo {author} {\bibfnamefont {T.}~\bibnamefont {Ðorđević}}, \bibinfo {author} {\bibfnamefont {P.}~\bibnamefont {Samutpraphoot}}, \bibinfo {author} {\bibfnamefont {P.~L.}\ \bibnamefont {Ocola}}, \bibinfo {author} {\bibfnamefont {H.}~\bibnamefont {Bernien}}, \bibinfo {author} {\bibfnamefont {B.}~\bibnamefont {Grinkemeyer}}, \bibinfo {author} {\bibfnamefont {I.}~\bibnamefont {Dimitrova}}, \bibinfo {author} {\bibfnamefont {V.}~\bibnamefont {Vuletić}},\ and\ \bibinfo {author} {\bibfnamefont {M.~D.}\ \bibnamefont {Lukin}},\ }\bibfield  {title} {\bibinfo {title} {Entanglement transport and a nanophotonic interface for atoms in optical tweezers},\ }\href {https://doi.org/10.1126/science.abi9917} {\bibfield  {journal} {\bibinfo  {journal} {Science}\ }\textbf {\bibinfo {volume} {373}},\ \bibinfo {pages} {1511} (\bibinfo {year} {2021})},\ \Eprint {https://arxiv.org/abs/https://www.science.org/doi/pdf/10.1126/science.abi9917} {https://www.science.org/doi/pdf/10.1126/science.abi9917}
  \BibitemShut {NoStop}%
\bibitem [{\citenamefont {Tanji-Suzuki}\ \emph {et~al.}(2011)\citenamefont {Tanji-Suzuki}, \citenamefont {Leroux}, \citenamefont {Schleier-Smith}, \citenamefont {Cetina}, \citenamefont {Grier}, \citenamefont {Simon},\ and\ \citenamefont {Vuletic}}]{Suzuki2011}%
  \BibitemOpen
  \bibfield  {author} {\bibinfo {author} {\bibfnamefont {H.}~\bibnamefont {Tanji-Suzuki}}, \bibinfo {author} {\bibfnamefont {I.~D.}\ \bibnamefont {Leroux}}, \bibinfo {author} {\bibfnamefont {M.~H.}\ \bibnamefont {Schleier-Smith}}, \bibinfo {author} {\bibfnamefont {M.}~\bibnamefont {Cetina}}, \bibinfo {author} {\bibfnamefont {A.~T.}\ \bibnamefont {Grier}}, \bibinfo {author} {\bibfnamefont {J.}~\bibnamefont {Simon}},\ and\ \bibinfo {author} {\bibfnamefont {V.}~\bibnamefont {Vuletic}},\ }\href@noop {} {\bibinfo {title} {Interaction between atomic ensembles and optical resonators: Classical description}} (\bibinfo {year} {2011}),\ \Eprint {https://arxiv.org/abs/1104.3594} {arXiv:1104.3594 [quant-ph]} \BibitemShut {NoStop}%
\bibitem [{\citenamefont {Carmichael}(1993)}]{carmichael_open_1993}%
  \BibitemOpen
  \bibfield  {author} {\bibinfo {author} {\bibfnamefont {H.}~\bibnamefont {Carmichael}},\ }\href@noop {} {\emph {\bibinfo {title} {An open systems approach to quantum optics: lectures presented at the {Université} {Libre} de {Bruxelles}, {October} 28 to {November} 4, 1991}}},\ \bibinfo {series} {Lecture notes in physics {New} series {M}, monographs}\ No.~\bibinfo {number} {18}\ (\bibinfo  {publisher} {Springer},\ \bibinfo {address} {Berlin Heidelberg},\ \bibinfo {year} {1993})\BibitemShut {NoStop}%
\bibitem [{\citenamefont {Steinberg}(2014)}]{steinberg_quantum_2014}%
  \BibitemOpen
  \bibfield  {author} {\bibinfo {author} {\bibfnamefont {A.}~\bibnamefont {Steinberg}},\ }\href {http://arxiv.org/abs/1406.5535} {\bibinfo {title} {Quantum {Measurements}: a modern view for quantum optics experimentalists}} (\bibinfo {year} {2014}),\ \bibinfo {note} {arXiv:1406.5535}\BibitemShut {NoStop}%
\bibitem [{\citenamefont {Omran}\ \emph {et~al.}(2019)\citenamefont {Omran}, \citenamefont {Levine}, \citenamefont {Keesling}, \citenamefont {Semeghini}, \citenamefont {Wang}, \citenamefont {Ebadi}, \citenamefont {Bernien}, \citenamefont {Zibrov}, \citenamefont {Pichler}, \citenamefont {Choi}, \citenamefont {Cui}, \citenamefont {Rossignolo}, \citenamefont {Rembold}, \citenamefont {Montangero}, \citenamefont {Calarco}, \citenamefont {Endres}, \citenamefont {Greiner}, \citenamefont {Vuletić},\ and\ \citenamefont {Lukin}}]{omran2019}%
  \BibitemOpen
  \bibfield  {author} {\bibinfo {author} {\bibfnamefont {A.}~\bibnamefont {Omran}}, \bibinfo {author} {\bibfnamefont {H.}~\bibnamefont {Levine}}, \bibinfo {author} {\bibfnamefont {A.}~\bibnamefont {Keesling}}, \bibinfo {author} {\bibfnamefont {G.}~\bibnamefont {Semeghini}}, \bibinfo {author} {\bibfnamefont {T.~T.}\ \bibnamefont {Wang}}, \bibinfo {author} {\bibfnamefont {S.}~\bibnamefont {Ebadi}}, \bibinfo {author} {\bibfnamefont {H.}~\bibnamefont {Bernien}}, \bibinfo {author} {\bibfnamefont {A.~S.}\ \bibnamefont {Zibrov}}, \bibinfo {author} {\bibfnamefont {H.}~\bibnamefont {Pichler}}, \bibinfo {author} {\bibfnamefont {S.}~\bibnamefont {Choi}}, \bibinfo {author} {\bibfnamefont {J.}~\bibnamefont {Cui}}, \bibinfo {author} {\bibfnamefont {M.}~\bibnamefont {Rossignolo}}, \bibinfo {author} {\bibfnamefont {P.}~\bibnamefont {Rembold}}, \bibinfo {author} {\bibfnamefont {S.}~\bibnamefont {Montangero}}, \bibinfo {author} {\bibfnamefont {T.}~\bibnamefont {Calarco}}, \bibinfo {author} {\bibfnamefont {M.}~\bibnamefont
  {Endres}}, \bibinfo {author} {\bibfnamefont {M.}~\bibnamefont {Greiner}}, \bibinfo {author} {\bibfnamefont {V.}~\bibnamefont {Vuletić}},\ and\ \bibinfo {author} {\bibfnamefont {M.~D.}\ \bibnamefont {Lukin}},\ }\bibfield  {title} {\bibinfo {title} {Generation and manipulation of schrödinger cat states in rydberg atom arrays},\ }\href {https://doi.org/10.1126/science.aax9743} {\bibfield  {journal} {\bibinfo  {journal} {Science}\ }\textbf {\bibinfo {volume} {365}},\ \bibinfo {pages} {570} (\bibinfo {year} {2019})}\BibitemShut {NoStop}%
\bibitem [{\citenamefont {Song}\ \emph {et~al.}(2019)\citenamefont {Song}, \citenamefont {Xu}, \citenamefont {Li}, \citenamefont {Zhang}, \citenamefont {Zhang}, \citenamefont {Liu}, \citenamefont {Guo}, \citenamefont {Wang}, \citenamefont {Ren}, \citenamefont {Hao}, \citenamefont {Feng}, \citenamefont {Fan}, \citenamefont {Zheng}, \citenamefont {Wang}, \citenamefont {Wang},\ and\ \citenamefont {Zhu}}]{song2019}%
  \BibitemOpen
  \bibfield  {author} {\bibinfo {author} {\bibfnamefont {C.}~\bibnamefont {Song}}, \bibinfo {author} {\bibfnamefont {K.}~\bibnamefont {Xu}}, \bibinfo {author} {\bibfnamefont {H.}~\bibnamefont {Li}}, \bibinfo {author} {\bibfnamefont {Y.-R.}\ \bibnamefont {Zhang}}, \bibinfo {author} {\bibfnamefont {X.}~\bibnamefont {Zhang}}, \bibinfo {author} {\bibfnamefont {W.}~\bibnamefont {Liu}}, \bibinfo {author} {\bibfnamefont {Q.}~\bibnamefont {Guo}}, \bibinfo {author} {\bibfnamefont {Z.}~\bibnamefont {Wang}}, \bibinfo {author} {\bibfnamefont {W.}~\bibnamefont {Ren}}, \bibinfo {author} {\bibfnamefont {J.}~\bibnamefont {Hao}}, \bibinfo {author} {\bibfnamefont {H.}~\bibnamefont {Feng}}, \bibinfo {author} {\bibfnamefont {H.}~\bibnamefont {Fan}}, \bibinfo {author} {\bibfnamefont {D.}~\bibnamefont {Zheng}}, \bibinfo {author} {\bibfnamefont {D.-W.}\ \bibnamefont {Wang}}, \bibinfo {author} {\bibfnamefont {H.}~\bibnamefont {Wang}},\ and\ \bibinfo {author} {\bibfnamefont {S.-Y.}\ \bibnamefont {Zhu}},\ }\bibfield  {title} {\bibinfo
  {title} {Generation of multicomponent atomic schrödinger cat states of up to 20 qubits},\ }\href {https://doi.org/10.1126/science.aay0600} {\bibfield  {journal} {\bibinfo  {journal} {Science}\ }\textbf {\bibinfo {volume} {365}},\ \bibinfo {pages} {574} (\bibinfo {year} {2019})},\ \Eprint {https://arxiv.org/abs/https://www.science.org/doi/pdf/10.1126/science.aay0600} {https://www.science.org/doi/pdf/10.1126/science.aay0600} \BibitemShut {NoStop}%
\bibitem [{\citenamefont {Moses}\ \emph {et~al.}(2023)\citenamefont {Moses} \emph {et~al.}}]{moses2023}%
  \BibitemOpen
  \bibfield  {author} {\bibinfo {author} {\bibfnamefont {S.~A.}\ \bibnamefont {Moses}} \emph {et~al.},\ }\bibfield  {title} {\bibinfo {title} {A race-track trapped-ion quantum processor},\ }\href {https://doi.org/10.1103/PhysRevX.13.041052} {\bibfield  {journal} {\bibinfo  {journal} {Phys. Rev. X}\ }\textbf {\bibinfo {volume} {13}},\ \bibinfo {pages} {041052} (\bibinfo {year} {2023})}\BibitemShut {NoStop}%
\bibitem [{\citenamefont {Degen}\ \emph {et~al.}(2017)\citenamefont {Degen}, \citenamefont {Reinhard},\ and\ \citenamefont {Cappellaro}}]{degen2017}%
  \BibitemOpen
  \bibfield  {author} {\bibinfo {author} {\bibfnamefont {C.~L.}\ \bibnamefont {Degen}}, \bibinfo {author} {\bibfnamefont {F.}~\bibnamefont {Reinhard}},\ and\ \bibinfo {author} {\bibfnamefont {P.}~\bibnamefont {Cappellaro}},\ }\bibfield  {title} {\bibinfo {title} {Quantum sensing},\ }\href {https://doi.org/10.1103/RevModPhys.89.035002} {\bibfield  {journal} {\bibinfo  {journal} {Rev. Mod. Phys.}\ }\textbf {\bibinfo {volume} {89}},\ \bibinfo {pages} {035002} (\bibinfo {year} {2017})}\BibitemShut {NoStop}%
\bibitem [{\citenamefont {Pezz\`e}\ \emph {et~al.}(2018)\citenamefont {Pezz\`e}, \citenamefont {Smerzi}, \citenamefont {Oberthaler}, \citenamefont {Schmied},\ and\ \citenamefont {Treutlein}}]{philipp2018}%
  \BibitemOpen
  \bibfield  {author} {\bibinfo {author} {\bibfnamefont {L.}~\bibnamefont {Pezz\`e}}, \bibinfo {author} {\bibfnamefont {A.}~\bibnamefont {Smerzi}}, \bibinfo {author} {\bibfnamefont {M.~K.}\ \bibnamefont {Oberthaler}}, \bibinfo {author} {\bibfnamefont {R.}~\bibnamefont {Schmied}},\ and\ \bibinfo {author} {\bibfnamefont {P.}~\bibnamefont {Treutlein}},\ }\bibfield  {title} {\bibinfo {title} {Quantum metrology with nonclassical states of atomic ensembles},\ }\href {https://doi.org/10.1103/RevModPhys.90.035005} {\bibfield  {journal} {\bibinfo  {journal} {Rev. Mod. Phys.}\ }\textbf {\bibinfo {volume} {90}},\ \bibinfo {pages} {035005} (\bibinfo {year} {2018})}\BibitemShut {NoStop}%
\bibitem [{\citenamefont {Li}\ \emph {et~al.}(2015)\citenamefont {Li}, \citenamefont {Humphreys}, \citenamefont {Mendoza},\ and\ \citenamefont {Benjamin}}]{Li2015}%
  \BibitemOpen
  \bibfield  {author} {\bibinfo {author} {\bibfnamefont {Y.}~\bibnamefont {Li}}, \bibinfo {author} {\bibfnamefont {P.~C.}\ \bibnamefont {Humphreys}}, \bibinfo {author} {\bibfnamefont {G.~J.}\ \bibnamefont {Mendoza}},\ and\ \bibinfo {author} {\bibfnamefont {S.~C.}\ \bibnamefont {Benjamin}},\ }\bibfield  {title} {\bibinfo {title} {Resource costs for fault-tolerant linear optical quantum computing},\ }\href {https://doi.org/10.1103/PhysRevX.5.041007} {\bibfield  {journal} {\bibinfo  {journal} {Phys. Rev. X}\ }\textbf {\bibinfo {volume} {5}},\ \bibinfo {pages} {041007} (\bibinfo {year} {2015})}\BibitemShut {NoStop}%
\bibitem [{\citenamefont {Kimble}(1998)}]{kimble1998}%
  \BibitemOpen
  \bibfield  {author} {\bibinfo {author} {\bibfnamefont {H.~J.}\ \bibnamefont {Kimble}},\ }\bibfield  {title} {\bibinfo {title} {Strong interactions of single atoms and photons in cavity qed},\ }\href {https://doi.org/10.1238/Physica.Topical.076a00127} {\bibfield  {journal} {\bibinfo  {journal} {Physica Scripta}\ }\textbf {\bibinfo {volume} {1998}},\ \bibinfo {pages} {127} (\bibinfo {year} {1998})}\BibitemShut {NoStop}%
\bibitem [{\citenamefont {Borregaard}\ \emph {et~al.}(2015)\citenamefont {Borregaard}, \citenamefont {K\'om\'ar}, \citenamefont {Kessler}, \citenamefont {S\o{}rensen},\ and\ \citenamefont {Lukin}}]{Borregaard2015}%
  \BibitemOpen
  \bibfield  {author} {\bibinfo {author} {\bibfnamefont {J.}~\bibnamefont {Borregaard}}, \bibinfo {author} {\bibfnamefont {P.}~\bibnamefont {K\'om\'ar}}, \bibinfo {author} {\bibfnamefont {E.~M.}\ \bibnamefont {Kessler}}, \bibinfo {author} {\bibfnamefont {A.~S.}\ \bibnamefont {S\o{}rensen}},\ and\ \bibinfo {author} {\bibfnamefont {M.~D.}\ \bibnamefont {Lukin}},\ }\bibfield  {title} {\bibinfo {title} {Heralded quantum gates with integrated error detection in optical cavities},\ }\href {https://doi.org/10.1103/PhysRevLett.114.110502} {\bibfield  {journal} {\bibinfo  {journal} {Phys. Rev. Lett.}\ }\textbf {\bibinfo {volume} {114}},\ \bibinfo {pages} {110502} (\bibinfo {year} {2015})}\BibitemShut {NoStop}%
\bibitem [{\citenamefont {Cohen-Tannoudji}\ \emph {et~al.}(1998)\citenamefont {Cohen-Tannoudji}, \citenamefont {Dupont-Roc},\ and\ \citenamefont {Grynberg}}]{cohen1998}%
  \BibitemOpen
  \bibfield  {author} {\bibinfo {author} {\bibfnamefont {C.}~\bibnamefont {Cohen-Tannoudji}}, \bibinfo {author} {\bibfnamefont {J.}~\bibnamefont {Dupont-Roc}},\ and\ \bibinfo {author} {\bibfnamefont {G.}~\bibnamefont {Grynberg}},\ }\href@noop {} {\emph {\bibinfo {title} {Atom-Photon Interactions}}}\ (\bibinfo  {publisher} {John Wiley and Sons},\ \bibinfo {year} {1998})\BibitemShut {NoStop}%
\bibitem [{\citenamefont {Scully}\ and\ \citenamefont {Zubairy}(1997)}]{scully1997}%
  \BibitemOpen
  \bibfield  {author} {\bibinfo {author} {\bibfnamefont {M.~O.}\ \bibnamefont {Scully}}\ and\ \bibinfo {author} {\bibfnamefont {M.~S.}\ \bibnamefont {Zubairy}},\ }\href {https://doi.org/10.1017/CBO9780511813993} {\emph {\bibinfo {title} {Quantum Optics}}}\ (\bibinfo  {publisher} {Cambridge University Press},\ \bibinfo {year} {1997})\BibitemShut {NoStop}%
\bibitem [{\citenamefont {Lukin}(2016)}]{lukin2016}%
  \BibitemOpen
  \bibfield  {author} {\bibinfo {author} {\bibfnamefont {M.~D.}\ \bibnamefont {Lukin}},\ }\href {https://lukin.physics.harvard.edu/files/lukin/files/physics_285b_lecture_notes.pdf} {\emph {\bibinfo {title} {Modern Atomic and Optical Physics II Lecture Notes}}}\ (\bibinfo {year} {2016})\BibitemShut {NoStop}%
\end{thebibliography}%

\end{document}


\preprint{APS/123-QED}

\title{Supplementary Material: Any-to-any connected cavity-mediated architecture for quantum computing with trapped ions or Rydberg arrays}

\author{Joshua Ramette}
\author{Josiah Sinclair}
\author{Zachary Vendeiro}
\author{Alyssa Rudelis}
\author{Marko Cetina}
\author{Vladan Vuleti\'c}

\date{\today}

\maketitle

\begin{align}
    \sum_m c_m \ket{m} \rightarrow \sum_m c_m e^{i \phi_m} e^{-\ell_m} \ket{m}
\end{align}

Argument for tuning up the phases and stabilization:

Equip each $\ket{m}$ state with two tones: one tone directly on resonance, and another tone slightly off resonance, that can be used to shift the energy and imprint a phase.

To deal with imperfections and nudge the many body state to the amplitudes and phases that you desire, imaging iterating with rounds of touch ups. First, start with the state:

and apply these beams to move you to the target amplitude and phase on each component. 

\begin{align}
    \sum_m c_m e^{i \phi_m} e^{-\ell_m} \ket{m}
\end{align}

The values $\phi_m, \ell_m$ for a given $m$ resulting from turning on a tone detuned by $\delta_m$ for time $t$ is:

$\phi_m = \frac{w^2 \delta_m}{\delta_m^2 + \kappa^2} t$

$\ell_m = \frac{w^2 \kappa}{\delta_m^2 + \kappa^2} t$

Therefore, when using a phase-imprint beam with for example $\delta_m \rightarrow \kappa$, we get $\phi_m \rightarrow \frac{w^2}{2 \kappa} t$, $\ell_m \rightarrow \phi_m t$. Therefore, the loss exponent is the same as the phase exponent. Since $\phi \leq \pi$, the loss factor is $\leq e^{-\pi} \approx 0.04$. Can make this a few times $\kappa$ for more favorable realistic scenario. For somewhat large $\delta_m = b \kappa$ would get $\ell_m \rightarrow \phi_m \frac{1}{b}$, so with $\delta_m = 4 \kappa$, loss factor would be $\leq e^{-\pi/4} \approx 0.46$ to imprint a phase of $\pi$.

Now we will assume $h \gg \kappa$, so we have well separated peaks. How much did the phase drives mess up the neighbors?:

a phase pulse on a level $m$ would cause phase and amp shifts on a neighbor $k$
$c_k \rightarrow c_k e^{i \phi_m \sqrt{N/C}/(k-m) } e^{-\phi_m (N/C)/(k-m)^2 }$

because we know $\sum_k^N \frac{1}{k} \leq (1 + \log(N)) \equiv A$ and $\sum_k^N \frac{1}{k^2} \leq \pi^2/6$, and for all $m$ $\phi_m < \phi_{max}$ we then know that $k$ is affected by all the other levels $m$ by at most:

$c_k \rightarrow c_k e^{i \phi_{max} A \sqrt{N/C} } e^{-\phi_{max} (N/C)(\pi^2/6) }$.


And the amplitude reduction drives also caused

neighbor $k$ to go from $c_k \rightarrow c_k e^{i \ell_m \sqrt{N/C}/(k-m)} e^{- \ell_m (N/C)/(k-m)^2}$

Then letting $\ell_{max}$ be the largest $\ell_m$, all of the $m$ amplitude reduction drives can have affected the level $k$ by at most
$c_k \rightarrow c_k e^{i \ell_{max} \sqrt{N/C} A} e^{- \ell_{max} (N/C) \pi^2/6}$

The many-body state is as desired up to these corrections. Now, we iterate again, applying local corrections to each $m$. First, we apply the phase shifting beams to cancel out the extra little phases. The amount of phase we have to cancel out, $\phi_{max}' \leq A \sqrt{N/C} (\phi_{max} + \ell_{max})$ can be made small for sufficiently large $C/N$. Furthermore, the amplitudes might have to also be reduced to push them back to the target amps across them all, up to the amount $e^{-\ell_{max}'} \equiv e^{-(N/C) \pi^2/6 (l_{max} + \phi_{max}) }$ for $\ell_{max}' \equiv (N/C) \pi^2/6 (l_{max} + \phi_{max})$. 

So long as $N/C$ is sufficiently small, $\phi_{max}, \ell_{max}$ will get multiplied by a factor smaller than one each iteration, converging exponentially to 0. The sum of all the $\ell_{max}$ across all the rounds is then a geometric series, so that the total loss would converge to something like $e^{-1/(1 - N/C (\pi^2/6) )}$ or maybe $e^{-1/(1 - \sqrt{N/C} A )}$. For small $\sqrt{N/C}A$, the loss is approximately $e^{-1/(1 - \sqrt{N/C} A )}$

neighboring amplitudes got extra decay of $e^{-d N/C}$ from the decay stuff. and got extra phase shift of $e^{-i d' \delta_0/h}$ from neighbors phase shifting.

Each term now must be iterated over, and have its amps and phase adjusted by these amounts. 

Rules for how a local phase and amp pulse affects neighbors:

adjust locally by $\phi$:

neighbors go from $c_m \rightarrow c_m e^{i \phi (\delta_0/h)} e^{-\phi (N/C)}$

adjust locally to reduce amplitude by $e^{-d}$:

neighbors go from $c_m \rightarrow c_m e^{i d \sqrt{N/C}} e^{-d (N/C)}$

After both rounds, we have a total phase distortion of $\phi (\delta_0/h) + d \sqrt{N/C} \equiv \phi' = (\phi + d) \times \sqrt{N/C}$.

and a total amp distortion of $e^{-\phi (N/C)} e^{-d (N/C)} \equiv e^{-d'}$ for $d' = \phi (N/C) + d (N/C) = (\phi + d) \times (N/C)$

Let G = $max(\phi_0, d_0)$

Then $\phi_1 = G \sqrt{N/C}$, $d_1 = G N/C$. Assuming $\sqrt{N/C}< 1$, subsequent corrections to $\phi, d$ are no larger than $\phi_1 \sqrt{N/C}$. Subsequent corrections then converge like $\phi_0 \times \sqrt{N/C}^j$ for $j$ rounds of corrections, converging to $\phi_0 \times \frac{1}{1 - \sqrt{N/C}}$

Now we can move on to the next round, where we have to apply local corrections of order $\phi', d'$.

Now assume $i d \sqrt{N/C} < 1$.

Each round, the amp and phase corrections gets smaller.


\bibliography{ramette_biblio}